\documentclass{aastex61}

\pdfoutput=1 
\usepackage{amsmath,amstext}
\usepackage{bold-extra}
\usepackage[figure,figure*]{hypcap}

\usepackage{graphicx} 
\usepackage{color}
\newcommand{\omm}{`Oumuamua}
\newcommand{\bldr}{BLDR 1.0}

\definecolor{black}{rgb}{0,0,0}
\definecolor{red}{rgb}{1.0,0,0}
\definecolor{blue}{rgb}{0,0,1.0}
\definecolor{green}{rgb}{0,1.0,0.0}
\definecolor{lightblue}{rgb}{0.4,0.8,1.0}
\definecolor{codeaccent}{rgb}{0.3,0.3,1.0}

\newcommand{\bb}[1]{\textbf{#1}}
\newcommand{\bc}[1]{\textcolor{codeaccent}{#1}}

\newcommand{\citePrice}{Price et al.~(submitted)}

\shorttitle{Breakthrough Listen: Data and Archiving}
\shortauthors{BL team}

\begin{document}

\title{The Breakthrough Listen Search for Intelligent Life: Public Data, Formats, Reduction and Archiving}

\newcommand{\UCB}{Department of Astronomy,  University of California Berkeley, Berkeley CA 94720}
\newcommand{\SSL}{Space Sciences Laboratory, University of California, Berkeley, Berkeley CA 94720}
\newcommand{\UCBEECS}{Department of Electrical Engineering and Computer Sciences,  University of California Berkeley, Berkeley CA 94720}
\newcommand{\SWIN}{Centre for Astrophysics \& Supercomputing, Swinburne University of Technology, Hawthorn, VIC 3122, Australia}
\newcommand{\GBT}{Green Bank Observatory,  West Virginia, 24944, USA}
\newcommand{\OXF}{Astronomy Department, University of Oxford, Keble Rd, Oxford, OX13RH, United Kingdom}
\newcommand{\NIJ}{Department of Astrophysics/IMAPP,Radboud University, Nijmegen, Netherlands}
\newcommand{\ATNF}{Australia Telescope National Facility, CSIRO, PO Box 76, Epping, NSW 1710, Australia}
\newcommand{\HOU}{Hellenic Open University, School of Science \& Technology, Parodos Aristotelous, Perivola Patron, Greece}
\newcommand{\USQ}{University of Southern Queensland, Toowoomba, QLD 4350, Australia}
\newcommand{\PENN}{Department of Astronomy and Astrophysics, Pennsylvania State University, University Park PA 16802}
\newcommand{\BTI}{Breakthrough Initiatives}
\newcommand{\SETI}{SETI Institute, Mountain View, California}
\newcommand{\KZA}{University of Malta, Institute of Space Sciences and Astronomy}
\newcommand{\ICRAR}{International Centre for Radio Astronomy Research, Curtin University, Australia}

\author{Matthew Lebofsky}
\affiliation{\UCB}

\author[0000-0003-4823-129X]{Steve Croft}
\affiliation{\UCB}

\author{Andrew P.\ V.\ Siemion}
\affiliation{\UCB}
\affiliation{\SETI}
\affiliation{\NIJ}
\affiliation{\KZA}

\author[0000-0003-2783-1608]{Danny C.\ Price}
\affiliation{\UCB}
\affiliation{\SWIN}

\author{J. Emilio Enriquez}
\affiliation{\UCB}
\affiliation{\NIJ}

\author[0000-0002-0531-1073]{Howard Isaacson}
\affiliation{\UCB}
\affiliation{\USQ}

\author{David H.\ E.\ MacMahon}
\affiliation{\UCB}

\author{David Anderson}
\affiliation{\SSL}

\author{Bryan Brzycki}
\affiliation{\UCB}

\author{Jeff Cobb}
\affiliation{\SSL}

\author[0000-0002-8071-6011]{Daniel Czech}
\affiliation{\UCB}

\author[0000-0003-3197-2294]{David DeBoer}
\affiliation{\UCB}

\author{Julia DeMarines}
\affiliation{\UCB}

\author{Jamie Drew}
\affiliation{\BTI}

\author{Griffin Foster}
\affiliation{\UCB}
\affiliation{\OXF}

\author[0000-0002-8604-106X]{Vishal Gajjar}
\affiliation{\UCB}

\author{Nectaria Gizani}
\affiliation{\UCB}
\affiliation{\HOU}

\author{Greg Hellbourg}
\affiliation{\ICRAR}

\author[0000-0001-8078-9395]{Eric J. Korpela}
\affiliation{\SSL}

\author{Brian Lacki}
\affiliation{Breakthrough Listen, \UCB}

\author{Sofia Sheikh}
\affiliation{\PENN}

\author{Dan Werthimer}
\affiliation{\UCB}

\author{Pete Worden}
\affiliation{\BTI}

\author{Alex Yu}
\affiliation{\UCBEECS}

\author{Yunfan Gerry Zhang}
\affiliation{\UCB}


\begin{abstract}
{\em Breakthrough Listen} is the most comprehensive and sensitive search for extraterrestrial intelligence (SETI) to date, employing a collection of international observational facilities including both radio and optical telescopes. During the first three years of the {\em Listen} program, thousands of targets have been observed with the Green Bank Telescope (GBT), Parkes Telescope and Automated Planet Finder. At GBT and Parkes, observations have been performed ranging from 700\,MHz to 26\,GHz, with raw data volumes averaging over 1\,PB / day. A pseudo-real time software spectroscopy suite is used to produce multi-resolution spectrograms amounting to approximately 400\,GB\,h$^{-1}$\,GHz$^{-1}$\,beam$^{-1}$. For certain targets, raw baseband voltage data is also preserved. Observations with the Automated Planet Finder produce both 2-dimensional and 1-dimensional high resolution (R $\sim 10^{5}$) echelle spectral data. 

Although the primary purpose of {\em Listen} data acquisition is for SETI, a range of secondary science has also been performed with these data, including studies of fast radio bursts. Other current and potential research topics include spectral line studies, searches for certain kinds of dark matter, probes of interstellar scattering, pulsar searches, radio transient searches and investigations of stellar activity. {\em Listen} data are also being used in the development of algorithms, including machine learning approaches to modulation scheme classification and outlier detection, that have wide applicability not just for astronomical research but for a broad range of science and engineering.

In this paper, we describe the hardware and software pipeline used for collection, reduction, archival, and public dissemination of {\em Listen} data. We describe the data formats and tools, and present Breakthrough Listen Data Release 1.0 (\bldr), a defined set of publicly-available raw and reduced data totalling 1\,PB.

\end{abstract}

\section{Introduction}

The Breakthrough Listen (hereafter BL) project \citep{doi:10.1089/space.2018.0027}, announced at the Royal Society in London in 2015 July, is currently undertaking searches for extraterrestrial intelligence (SETI) at telescope facilities across the globe. Motivated by the idea that some human technologies used for communication, propulsion, or other purposes, generate signatures that could be detected by a sufficiently sensitive telescope at interstellar distances, BL conducts observational astronomy programs targeting such ``technosignatures'' that may betray the presence of technology, and hence advanced life, on worlds other than our own.

Currently, BL undertakes the search for technosignatures at both radio and optical wavelengths. The radio search currently employs two large single-dish telescopes, the Robert C. Byrd Green Bank Telescope (GBT; \citealt{gbtinstrument}) in West Virginia, USA, and the CSIRO Parkes Radio Telescope \citep{parkesinstrument} in New South Wales, Australia. The optical search currently employs the Automated Planet Finder (APF) at Lick Observatory in California, USA. The program will soon expand to include radio observations using the MeerKAT telescope in South Africa and the VERITAS Cherenkov Telescope at the Whipple Observatory in Arizona, USA. Pilot programs and collaboration agreements are also in place for additional facilities, including the FAST 500m telescope in Guizhou Province, China \citep{Initiatives:2016up}, the Murchison Widefield Array in Western Australia \citep{2018ApJ...856...31T}, and the Jodrell Bank Observatory in Manchester, UK.

Core observational goals for the BL program include multiwavelength observations of $\sim$1700 nearby stars and 100 nearby galaxies, a radio survey of one million additional nearby stars and radio and optical observations of the Galactic Plane and the Galactic Center (GP/GC). The one million star survey will be performed using dedicated digital beamformer hardware at MeerKAT, allowing dozens of stars within the primary MeerKAT field of view to be observed simultaneously. These commensal (contemporaneously using the same data stream) observations will occur with the BL instrument alongside all other uses of the facility. 

In addition to the core program, BL occasionally observes other targets of interest. These have thus far included objects in our own Solar System, including the asteroid BZ\,509 \citep{2019RNAAS...3a..19P} the interstellar object \omm\ \citep{2018RNAAS...2a...9E}, as well as other ``exotica'' such as Boyajian's Star \citep{Wright2016}, Ross\,128 \citep{Enriquez2019}, candidate technosigntures from other research programs  \citep{Croft:2016tm} and a number of fast radio bursts (FRBs). The BL backends at GBT and Parkes have also been made available in shared-risk mode for proposals from the astronomical community, which are considered for allocations of general observer time by the telescope time allocation committees.

The bulk of observations conducted during the first three years of the BL program have focused on nearby stars drawn from the sample described by \citet{Isaacson2017}. An initial SETI search of GBT data on 692 stars from this sample was presented by \citet{enriquez2017}. Since 2016, additional stars from \citet{Isaacson2017} have been observed by the BL facilities, and our analysis techniques have been, and continue to be, refined. An updated analysis of radio observations of a larger sample of these stars is presented in \citePrice{}. 

This paper describes our data acquisition, formats, reduction, and archiving. We also announce the release of $\sim 1$\,PB of public data, which forms BL Data Release 1.0 (\bldr), and consists of:

\begin{itemize}
    \item Radio spectrograms for the 1327\ stars presented in the analysis of \citePrice{} -- 882\ from $L$-band and 1005\ from $S$-band receivers at GBT (Table \ref{tab:bands}) and 189\ from the  10-cm receiver at Parkes, along with corresponding observations of pulsar, flux calibrator, and reference (``off") sources
    \item Optical spectra, including both 2D and corresponding extracted 1D spectra, for 789 stars observed by APF
    \item Raw time domain voltage data and corresponding spectrograms from radio observations of FRB\,121102 (GBT) and FRB\,180301 (Parkes)
\end{itemize}

The remainder of this paper is structured as follows: Section~\ref{gbtparkes} describes data acquisition and reduction procedures for BL observations with GBT and Parkes, Section~\ref{apf} describes data acquisition and reduction procedures for BL observations with APF, and Section~\ref{archive} details the design and interface for the BL public data archive.  Appendix~\ref{sec:filenaming} describes the file naming conventions used for BL data products. Appendix~\ref{sec:fileheaders} lists the header information for raw and filterbank files. 
\\
\section{Breakthrough Listen on GBT and Parkes}\label{gbtparkes}

The digital backends used by BL at radio observatories consist of clusters of compute nodes with large amounts of storage, each with one or more graphics processing units (GPUs) for reduction and processing; storage nodes, with additional capacity intended for longer-term data storage, and head nodes that provide boot images and a login gateway to the compute and storage nodes. The GBT BL instrument \citep{gbtinstrument} saw first light in 2015 December, and currently consists of 9 storage nodes and 65 compute nodes (64 active plus one spare), with a total usable capacity of $\sim 8$\,PB. The Parkes BL instrument \citep{parkesinstrument} saw first light in 2016 September, and currently consists of 6 storage nodes and 27 compute nodes (26 active plus one spare), with capacity $3.5$\,PB. The upcoming MeerKAT instrument will be described in a future paper.

\begin{figure*}[h]
\centering
\includegraphics[width=0.9\linewidth]{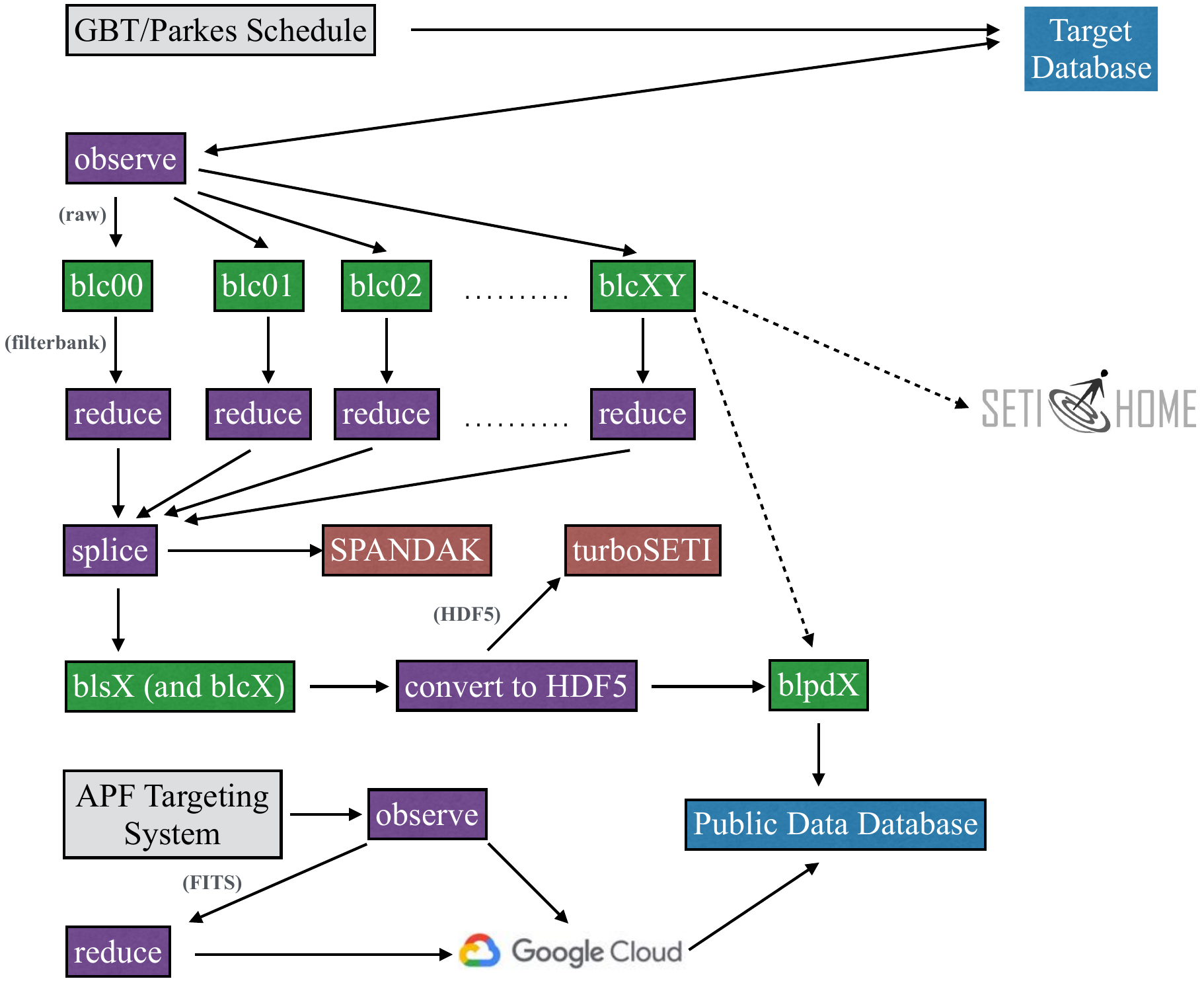}
\caption{\label{fig:pipeline}Flowchart illustrating the BL data pipeline. Purple-colored boxes represent the flow of data through the acquisition and reduction steps and brick-colored boxes represent analysis pipelines. Green boxes represent physical compute hardware and resources, i.e.\ the BL compute nodes (blcXY), storage nodes (blsX), public data nodes (blpdX), and cloud storage through which the data passes. Parenthetical tags show data formats when files are created or converted.  The dashed arrows illustrate that while most raw data are not archived indefinitely (\S~\ref{gbtparkes}), a small fraction of the raw data is made available as part of our public datasets, and during reduction a small fraction is passed to the SETI@home volunteer distributed computing project \citep{Anderson:2002:SEP:581571.581573}.).}
\end{figure*}

The basic data pipeline is illustrated graphically in Figure~\ref{fig:pipeline}, and is described in detail below. At Parkes and GBT, BL compute nodes receive digitized complex voltages from FPGA data acquisition systems \citep{gbtinstrument}, which are subsequently written to large disk arrays in GUPPI (Green Bank Ultimate Pulsar Processing Instrument; \citealt{guppi}) raw format.  The GUPPI format is derived from the FITS standard (\citealt{fits,fits3}), and has been used in a number of other SETI experiments that employed the original GUPPI instrument, beginning with \cite{2013ApJ...767...94S} and continuing more recently with \cite{Margot:2018} and \cite{Pinchuk:2019}. 

Each BL compute node (denoted blc00 through blcXY in Figure~\ref{fig:pipeline}) captures a portion of the total bandwidth during observing with a maximum of 187.5\,MHz of dual-polarization, 8-bit sampled voltage data at a rate of 6\,Gb\,s$^{-1}$ per node. While the raw data format allows for frequency overlap between nodes, BL does not implement this functionality. 

BL observing sessions last between 2 and 12 hours. During the gaps between sessions,which are at least as long as the sessions themselves, raw data are processed using a GPU-accelerated Fast Fourier Transform (FFT) code, and the resultant spectrograms are archived. For certain targets of interest, we also choose to archive some of the raw voltages to permit additional post-processing. One example of such a target is FRB\,121102; the 400\,TB raw dataset that was used in the analysis of \citet{gajjar:18} and \citet{zhang:18} is being made publicly available as part of \bldr.

Searches for technosignatures are performed with custom pipelines. These include 
 {\sc SPANDAK} \citep{gajjar:18}, used to search for broadband pulses, and {\sc TurboSETI} \citep{enriquez2017}, used to search for narrowband drifting signals, in addition to algorithms under development including machine learning approaches. 

\subsection{Observing strategy}

\label{sec:observingstrategy}
The standard BL observing procedure at GBT and Parkes is to observe in cadences of 30 minutes. Primary targets are selected from a MySQL target database (Figure~\ref{fig:pipeline}), and observing scripts are generated for each session based on designated target priorities and the amount of data already obtained for each target. The database is updated after each run. The observing script controls the observations via an interface with the facility software ({\sc astrid} and {\sc cleo} at GBT\footnote{GBT Observer's Guide - \url{https://science.nrao.edu/facilities/gbt/observing/GBTog.pdf}}; {\sc TCS} at Parkes\footnote{Parkes Radio Telescope Users Guide - \url{http://www.parkes.atnf.csiro.au/observing/documentation/user_guide/pks_ug.pdf}}), and also manages the operation of the BL backend cluster which records the digitized raw voltages from the telescope.

For single-pixel instruments at GBT and Parkes, an observing strategy of alternating (or ``nodding") between intended sources and nearby reference points in the sky is adopted in order to discriminate against the main contaminant for any radio SETI search: human generated radio frequency interference (RFI). The intensity and morphology of detected RFI signals may vary with time and with telescope position, but almost never in such a way that only the intended sources are affected but not the reference sources. By looking for signals detected in only the primary location, we thereby apply a spatial filter that preferentially isolates signals that may be coming from the direction of the primary target being observed.

Each primary target is observed for three 5-minute integrations, interspersed with three 5-minute scans of nearby comparison targets. Primary targets are denoted as ``A'' or ``ON'' targets. Early in the BL program our ``OFF'' targets were a nearby blank sky pointing (reflected with an ``\_OFF" suffix in the target name), but several months after our GBT program began we modified the strategy. Instead of nodding to a nearby blank part of the sky we now alternate the primary ``A'' targets with three different nearby stars (denoted ``B'', ``C'', and ``D''), such that our standard cadence is now ``ABACAD''. This method maximizes the number of observed stars while still allowing for RFI detection and removal. An example set of recently recorded Green Bank data is shown in Figure~\ref{fig:filterbank}.  Information regarding the file naming conventions is given in Appendix~\ref{sec:filenaming}.

\begin{figure*}
\centering
\includegraphics[width=0.35\linewidth]{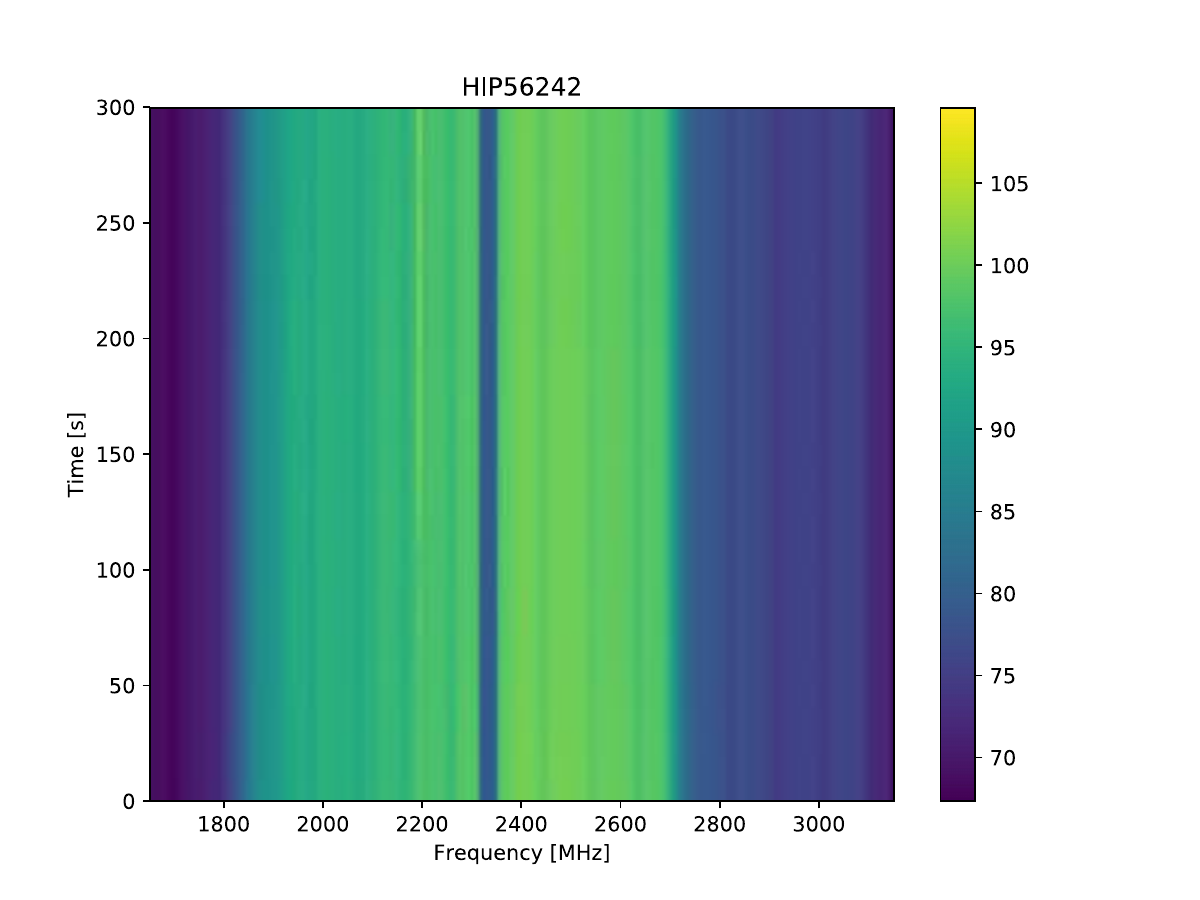}%
\includegraphics[width=0.35\linewidth]{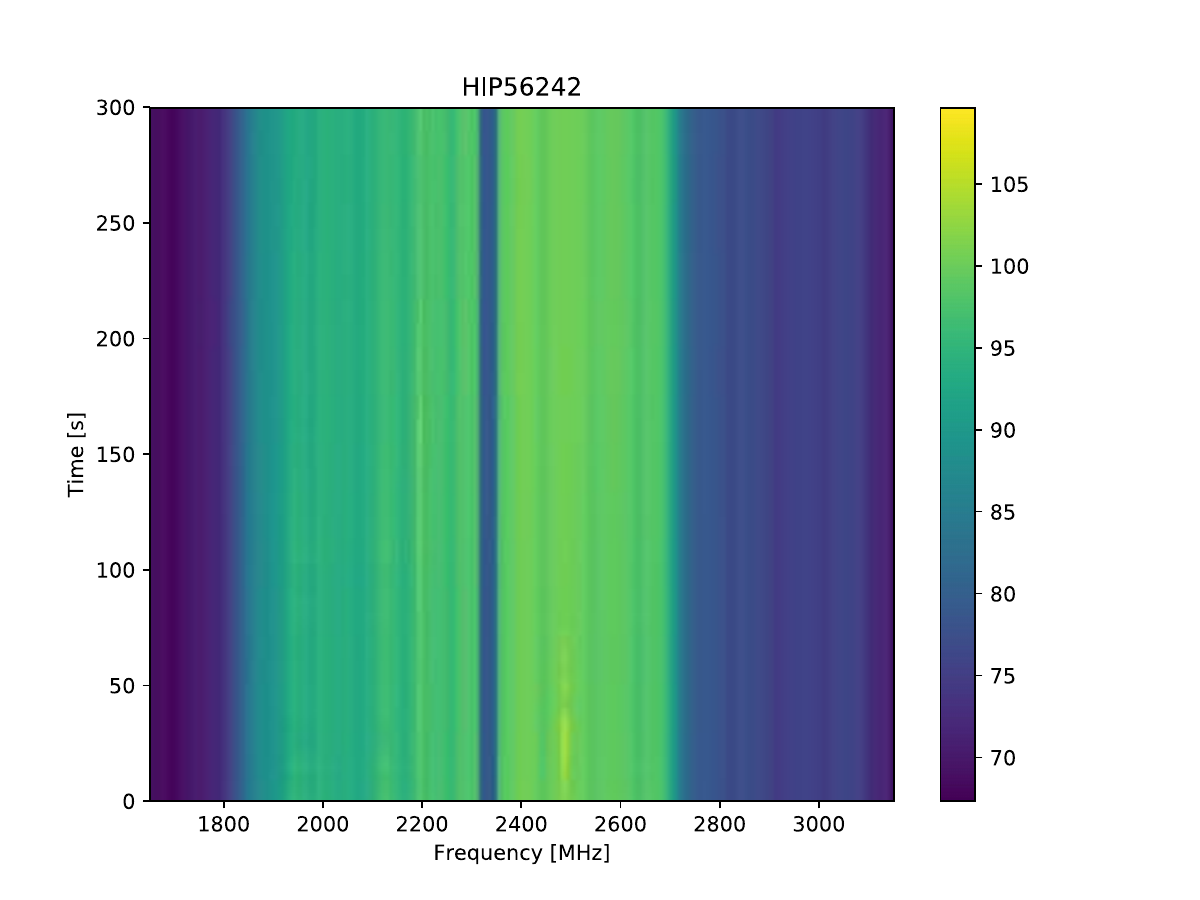}%
\includegraphics[width=0.35\linewidth]{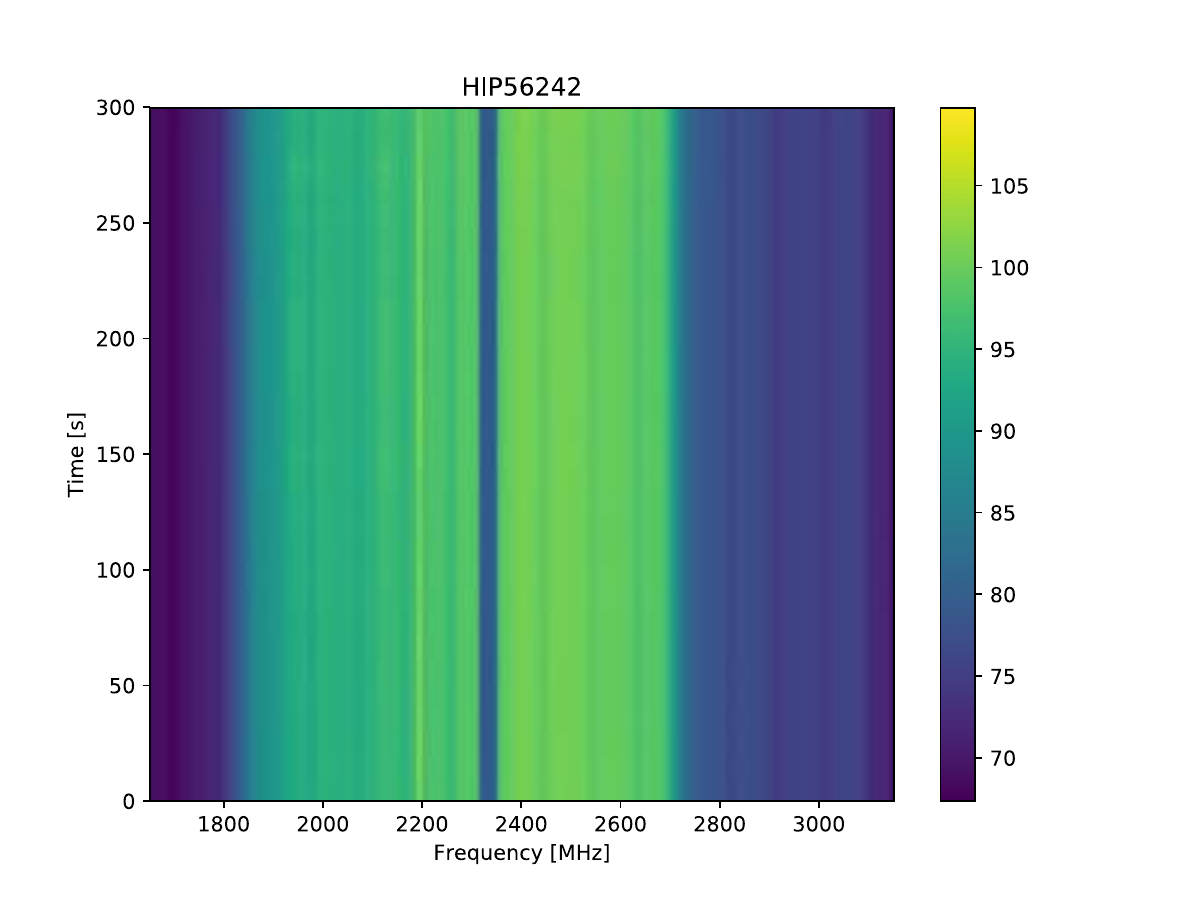}
\includegraphics[width=0.35\linewidth]{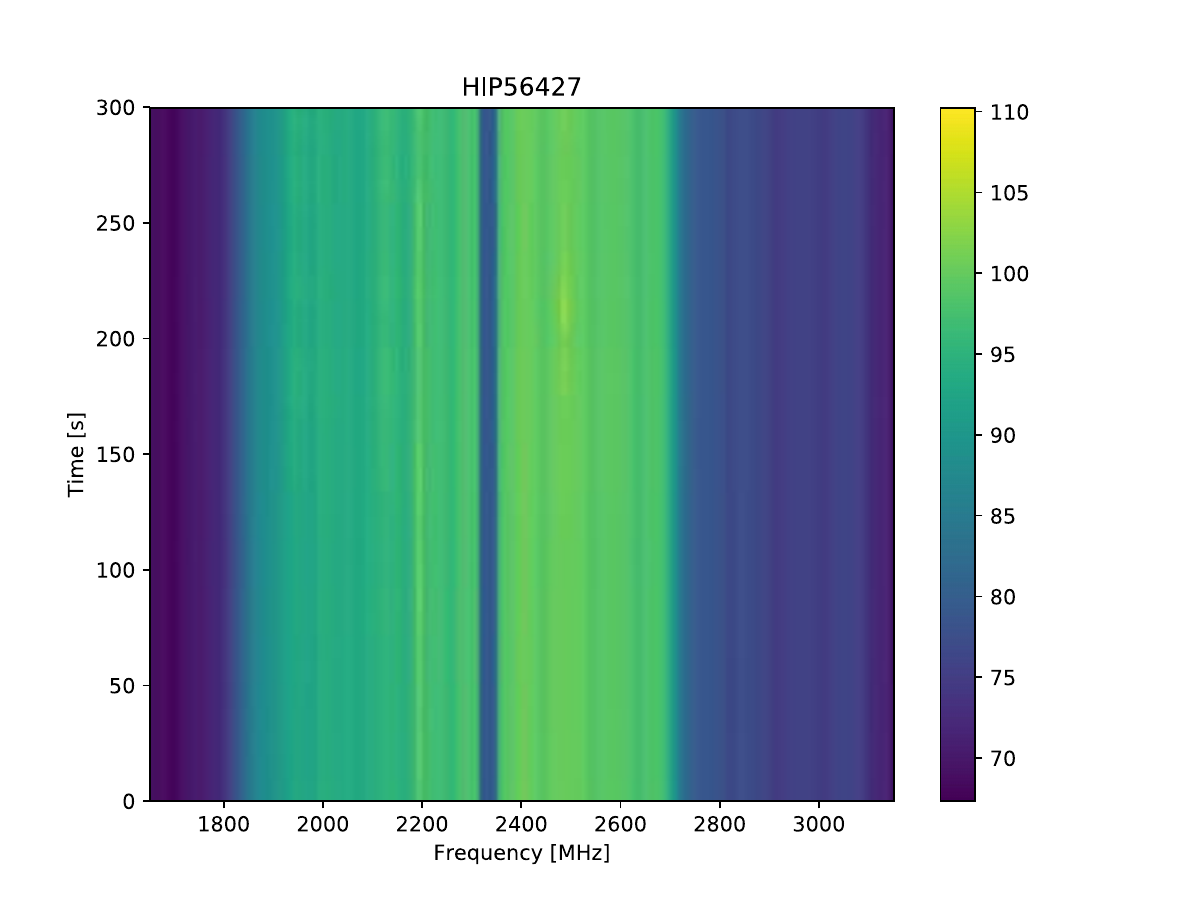}%
\includegraphics[width=0.35\linewidth]{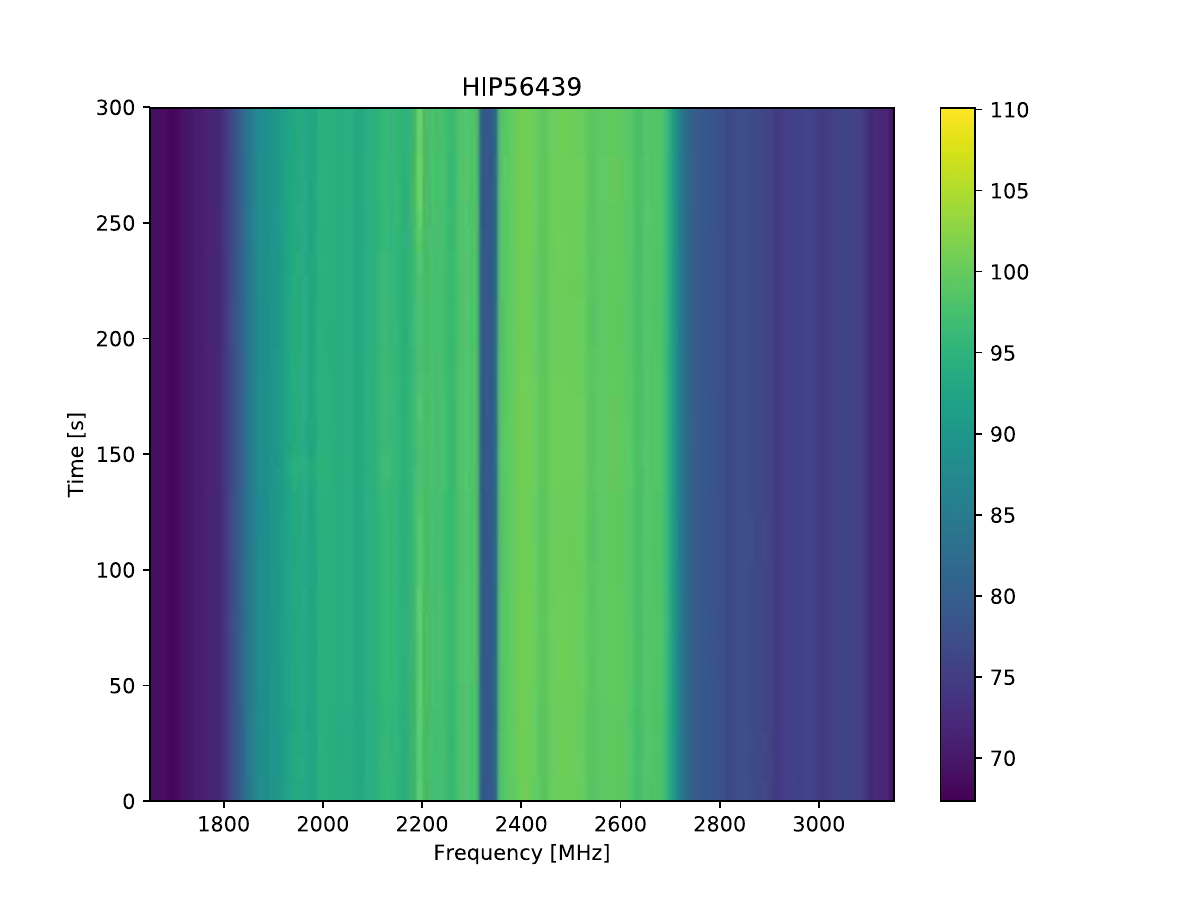}%
\includegraphics[width=0.35\linewidth]{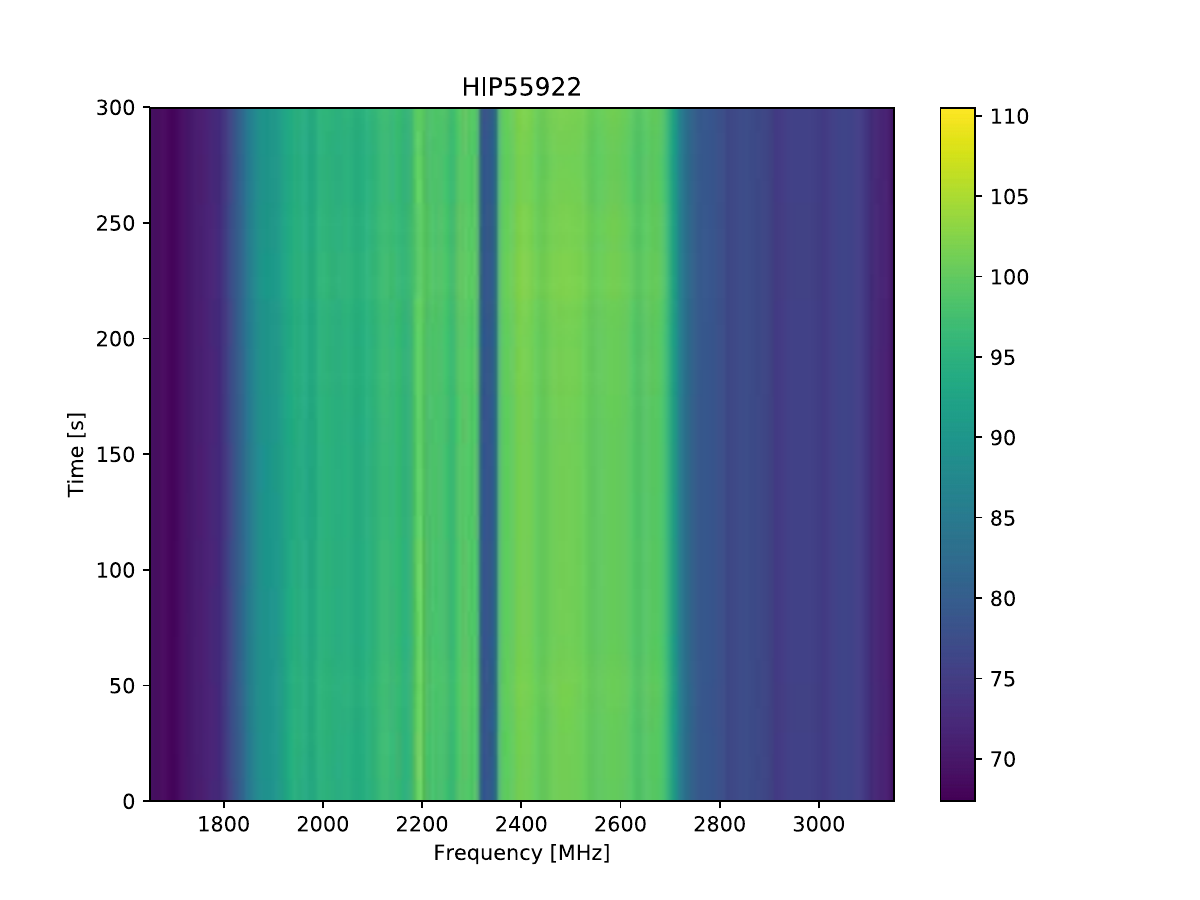}
\caption{\label{fig:filterbank}Spectrograms (time versus frequency) from an example ``ABACAD'' cadence showing S-band data spliced from 8 compute nodes, plotted from the corresponding filterbank (\S~\ref{sec:filterbank}) files using {\sc blimpy}. Observations of the primary (``A'') target (HIP\,56242 in this case) are alternated with the ``B'', ``C'', and ``D'' targets (in this case, HIP\,56427, HIP\,56439, and HIP\,55922, respectively). Good candidate technosignatures would be expected to appear only in observations of the ``A'' target (top row), and be absent from the comparison observations (bottom row).}
\end{figure*}

For Parkes we kept the original strategy of nodding between the target source and a nearby blank point in the sky (\citePrice). The target naming of on/off sources is different than at GBT due to {\sc TCS} conventions. The ``ON" targets have an ``\_S" suffix (source) and the ``OFF" targets have an ``\_R" suffix (reference). 

\begin{figure}
    \centering
    \includegraphics[scale=1]{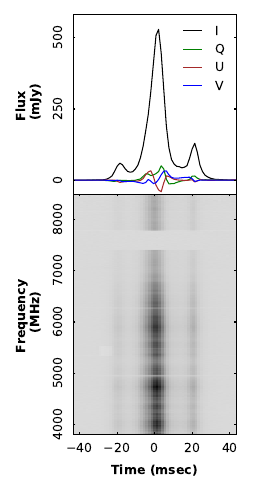}
    \caption{Detection of PSR B0329+54 across 4 to 8\,GHz using the BL backend at GBT, plotted using the PSRCHIVE software package. The top panel shows integrated full-Stokes profiles obtained from 5-minute long observations. The bottom panel shows the intensity as a function of frequency and pulse phase. Frequencies between 7100 to 7500\,MHz were flagged due to RFI. 
    }
    \label{fig:0329}
\end{figure}

We typically observe one bright pulsar and one bright radio calibrator at the beginning of each session. 
Immediately after each pulsar observation, the pulsar data are processed using the standard pulsar tools. We use {\sc prepfold} from PRESTO \citep{presto}, DSPSR \citep{2011PASA...28....1V}, and PSRCHIVE \citep{2012AR&T....9..237V} to check that the pulsar is detected, and as a diagnostic of instrument performance. The expected signal-to-noise ratio (SNR) is compared with the observed SNR as a diagnostic of instrument sensitivity 
(Figure~\ref{fig:0329}). Since pulsars are known to show scintillation prohibiting their use for flux calibration, we also observe standard flux calibrator sources during each session. Flux calibration is not currently a part of our standard reduction pipeline -- we perform detection and signal characterization in instrumental units, or perform an approximate calibration using the radiometer equation. However, work is underway to flux- and polarization-calibrate all of our data products. This work will be extended back to data in our archive using the calibrator sources that have been observed thus far.

\begin{figure}
    \centering
    \includegraphics[scale=0.55]{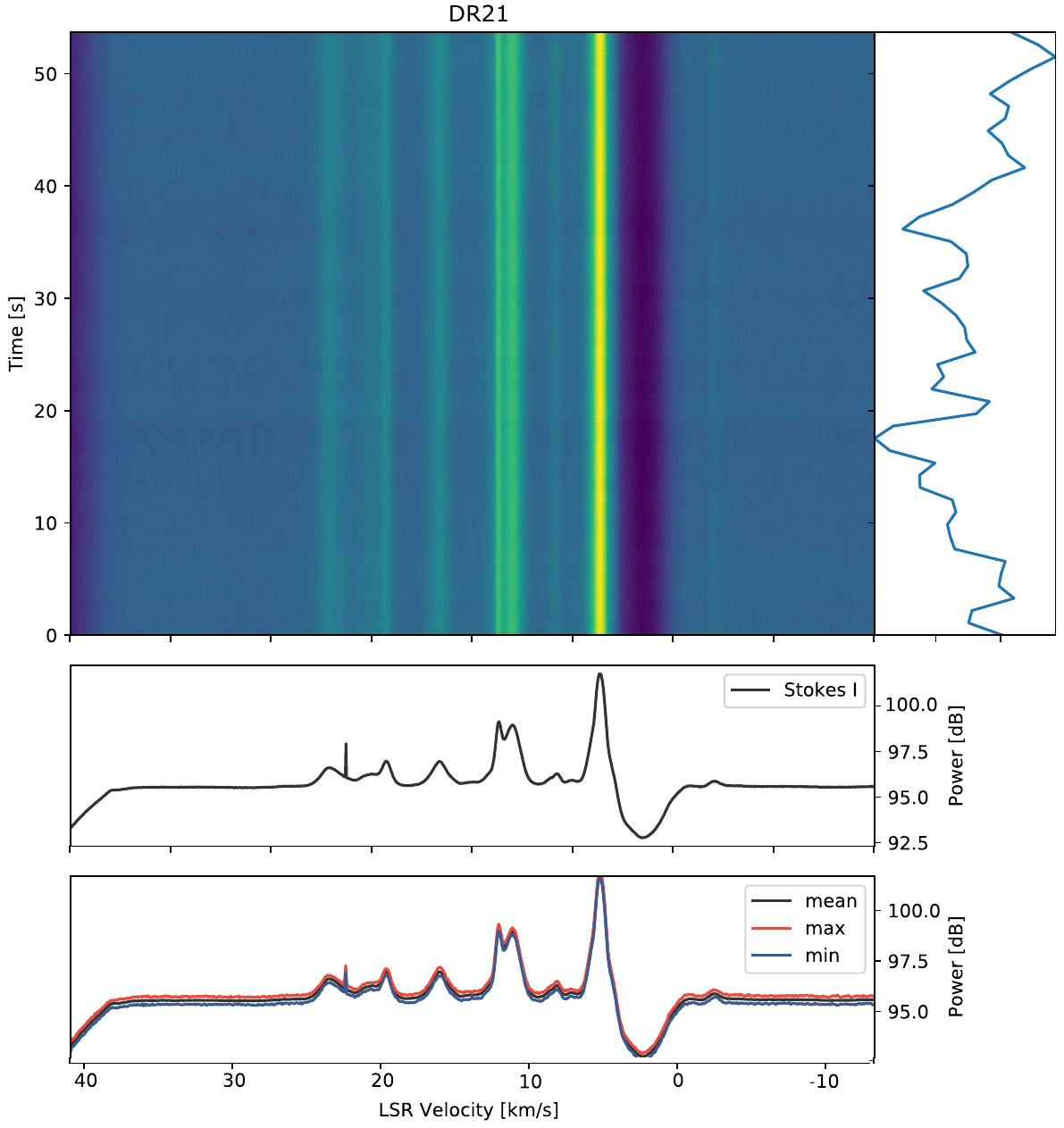}
    \caption{Detection of the 22\,GHz H$_{2}$O maser feature in DR\,21 using the KFPA receiver at GBT and the medium resolution (MR) BL data product as described in Table~\ref{tab:dataproducts}. he polyphase channel shape and DC feature described in Section \ref{sec:artifacts} are apparent. While we haven't made any K-band data available in this data release, this analysis is from data collected during early commissioning of this receiver and helps to further demonstrate our data and analysis potential even at this lower resolution.}
    \label{fig:maser}
\end{figure}

\begin{figure}
    \centering
    \includegraphics[scale=0.55]{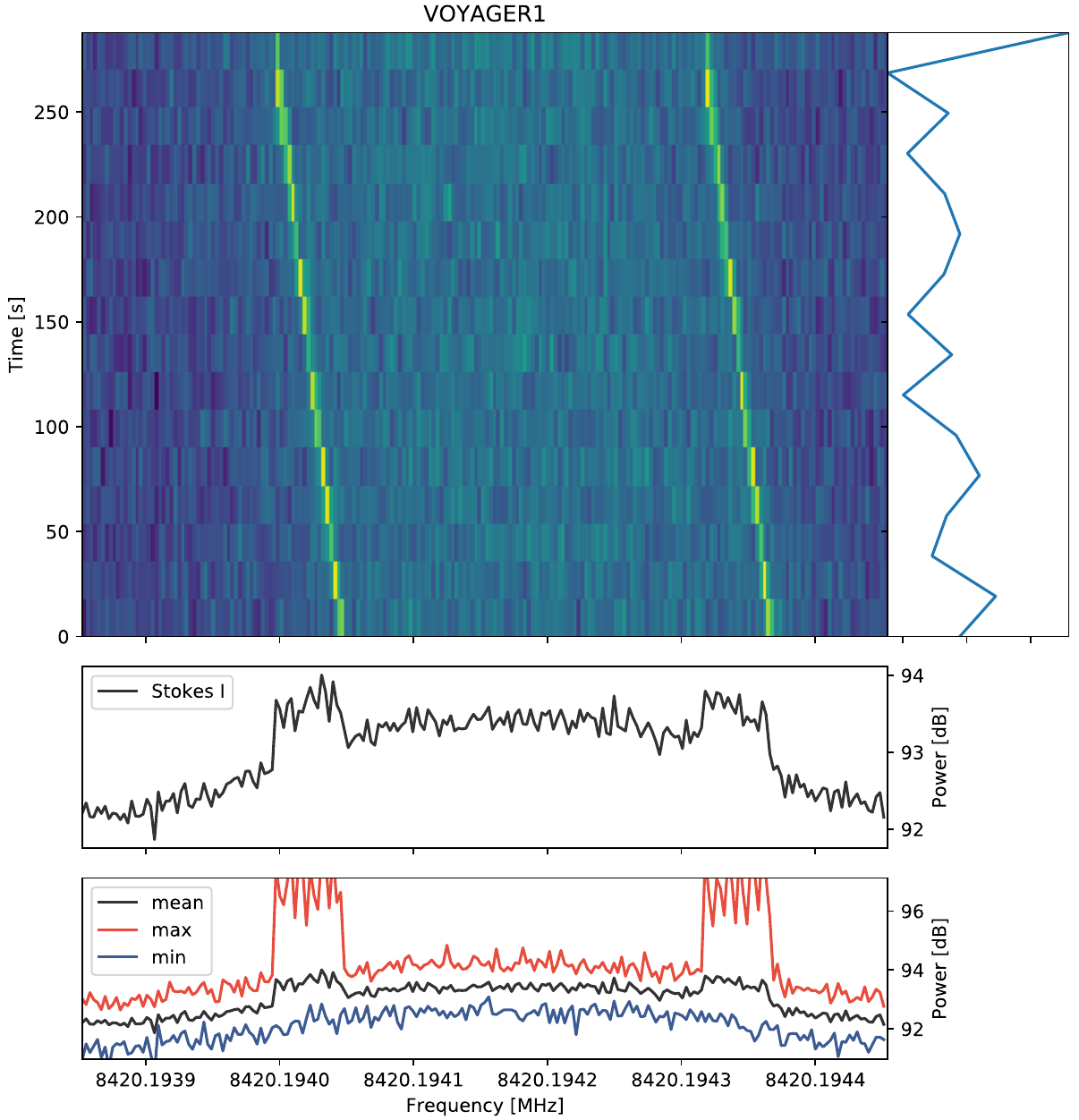}
    \caption{Detection of one sideband of the communication downlink from the spacecraft Voyager I using the X-band receiver at GBT and the high spectral resolution (HSR) BL data product as described in Table \ref{tab:dataproducts}. These data are presented in the topocentric frame for MJD 57386.8438889.}
    \label{fig:voyager}
\end{figure}

\begin{figure}
    \centering
    \includegraphics[scale=0.55]{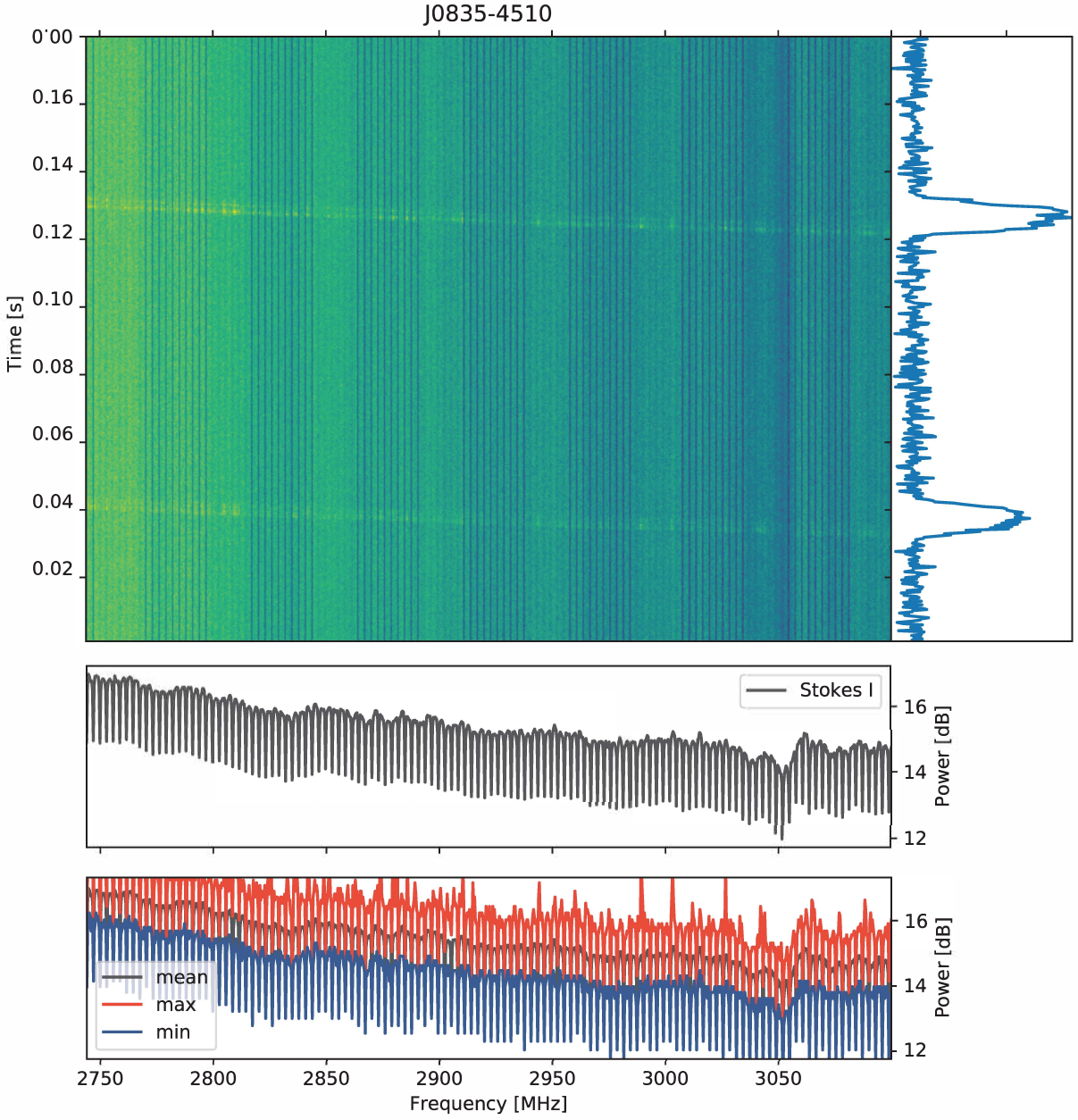}
    \caption{Detection of two pulses from the Vela Pulsar (J0835-4510) using the 1050 receiver at Parkes and the high time resolution (HTR) BL data product as described in Table \ref{tab:dataproducts}. These data are presented in the topocentric frame for MJD 58008.024838.}
    \label{fig:vela}
\end{figure}

At GBT we can record up to a 10\,GHz-wide band from any of the GBT receivers; this is the full bandwidth that the analog downconversion systems can deliver from any given receiver. The first three years of our GBT program have primarily consisted of stars drawn from the \citet{Isaacson2017} sample, observed at $L$-, $S$-, $C$-, and $X$-bands (Table~\ref{tab:bands}). The majority of the stars from that sample that are visible from GBT have now been observed at both $L$- and $S$-band (center frequencies of 1500 and 2400\,MHz respectively), and the associated data files are being made available in \bldr. Raw data files are native linear polarization; filterbank files are Stokes I.

\begin{deluxetable}{cccc}
\tablewidth{0pt}
\tabletypesize{\scriptsize}
\tablecaption{\label{tab:bands} Receivers currently in regular use for BL observations. }
\tablehead {
\colhead{Telescope} &
\colhead{Receiver} &
\colhead{Standard Center Frequency (MHz)} &
\colhead{Frequency Range for Public HDF5 Products}
}
\startdata
GBT & L-band & 1500 & 1025--1925 \\
GBT & S-band & 2400 & 1820--2720 \\
GBT & C-band & 6000 & TBD\\
GBT & X-band & 9375 & TBD\\
Parkes & 1050CM & 2900 & 2574--3444 \\
Parkes & Multibeam & 1382 & TBD\\
\enddata
\vspace{10pt}
Note: There are no C-band, X-band, or Parkes Multibeam HDF5 products as part of this release - we haven't yet decided how much we want to trim down these products and will make such determinations when we have a more complete data set.
\end{deluxetable}

At Parkes, we can record up to an aggregate 4.875\,GHz of bandwidth, dual polarization (native linear), either from single-pixel feeds such as the 10-cm receiver or the Ultra-Wide-Band feed (Hobbs et. al., in prep), or distributed as 308\,MHz bands from each of the 13 beams of the Multibeam receiver. We focused on nearby stars with declinations less than $-20$ degrees from the sample of \cite{Isaacson2017} with the 10-cm receiver, and the associated data files are being made available in \bldr. 

We are in process of commissioning use of additional receivers at both telescopes -- for example the K-band focal plane array (KFPA) at GBT, and the Ultra-wideband receiver (UWL) at Parkes. Data from these receivers will be made available in the future.

\subsection{Raw data}

Each raw file consists of a sequence of header units and corresponding chunks of data referred to as blocks, representing data collected over the course of an observation in time-contiguous order. 

Each line in the headers conforms to the FITS standard of 80 bytes (an 8 byte keyword and associated values). To reduce overhead and maximize the rate at which data are written to disk, files are written using Direct I/O, a method by which applications can bypass the operating system caches and write more directly to storage devices. This mode is faster and preferred when doing long sequential writes of random data (and thus caching is ineffective). However Direct I/O requires 512 byte writes. Headers must therefore be padded with zeros after the last 80 byte FITS header to round them up multiples of 512 bytes. This padding is incompatible with many existing data reduction tools which expect the data payload to begin immediately after the header's last key/value pair. At first our pipeline contained a ``fix up'' script, rewriting the files without the header padding. We later updated the reduction tools to handle raw data files with or without this padding. The two methods are differentiated by an extra field in the header (keyword {\sc directio}) which is either 0 to reflect that the current header is not padded or 1 to reflect the header is padded. The last line contains {\sc end}, after which the binary data, or the padding up to the next 512 byte boundary, begins.

Raw digitized voltages are coarsely channelized into 512 polyphase channels \citep{gbtinstrument,parkesinstrument} across a Nyquist bandwidth of 1500\,MHz, resulting in 64 channels over 187.5\,MHz per compute node and a coarse channel width $\mathrm{BW}_{\rm{polyphase}} = 2.93$\,MHz. For L- and S-band observations, the RF band is inverted at IF, resulting in an RF band ranging from $f_{\mathrm {tuning}} + 750$\,MHz to $ f_{\mathrm {tuning}} - 750$\,MHz in the IF range $0 - 1500$\,MHz. Frequency inversion can be deduced from the FITS keyword {\sc OBSBW}. The absolute value represents the bandwidth of the file in MHz; if it is negative the frequencies are inverted. 

\begin{figure*}
\centering
\includegraphics[width=\linewidth]{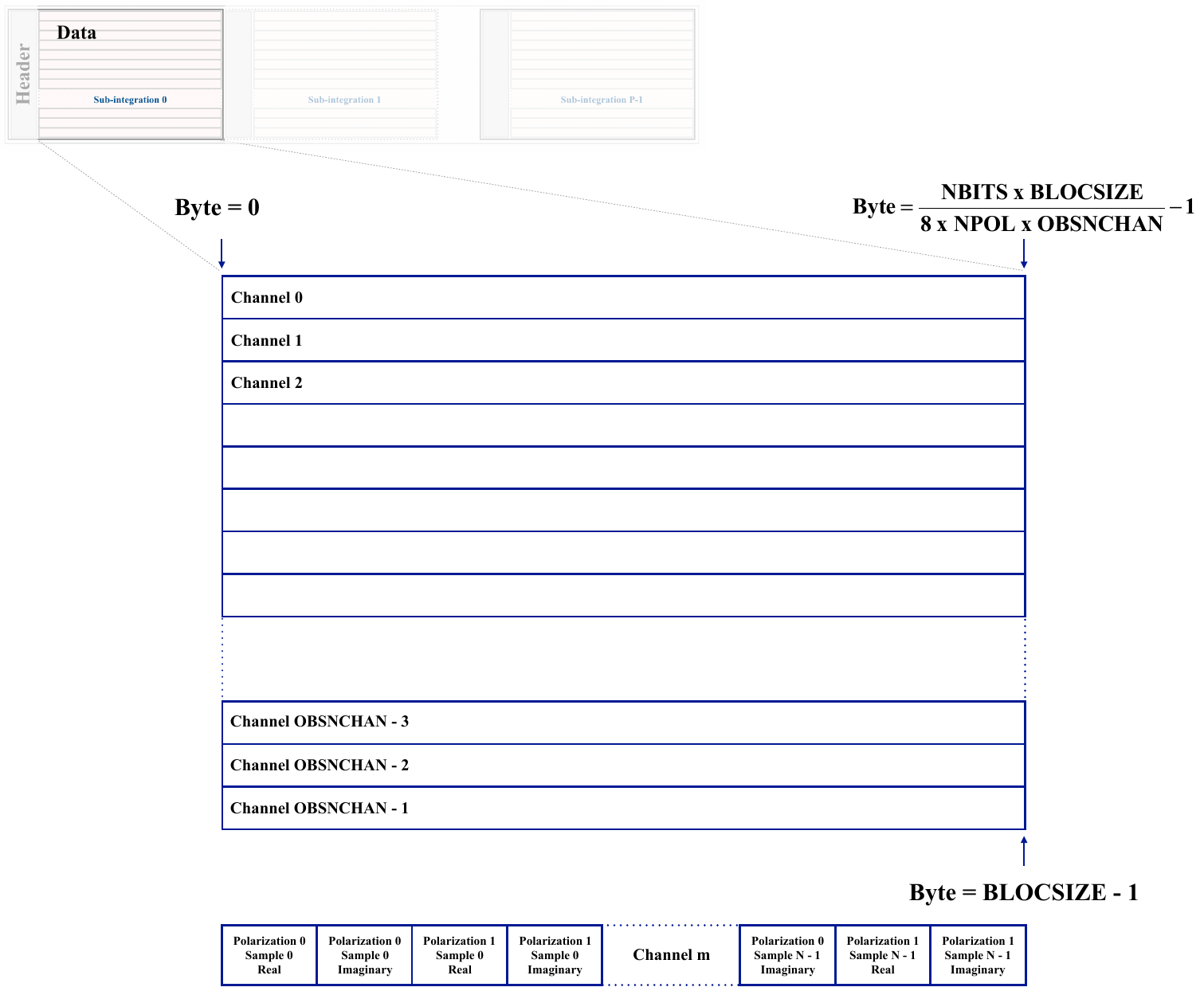}
\caption{\label{fig:rawchannels} GUPPI raw format showing frequency channelization of each sub-integration. The values used in byte calculations are raw data header keywords described in Appendix~\ref{sec:fileheaders}.}
\end{figure*}

\begin{figure*}
\centering
\includegraphics[width=0.8\linewidth]{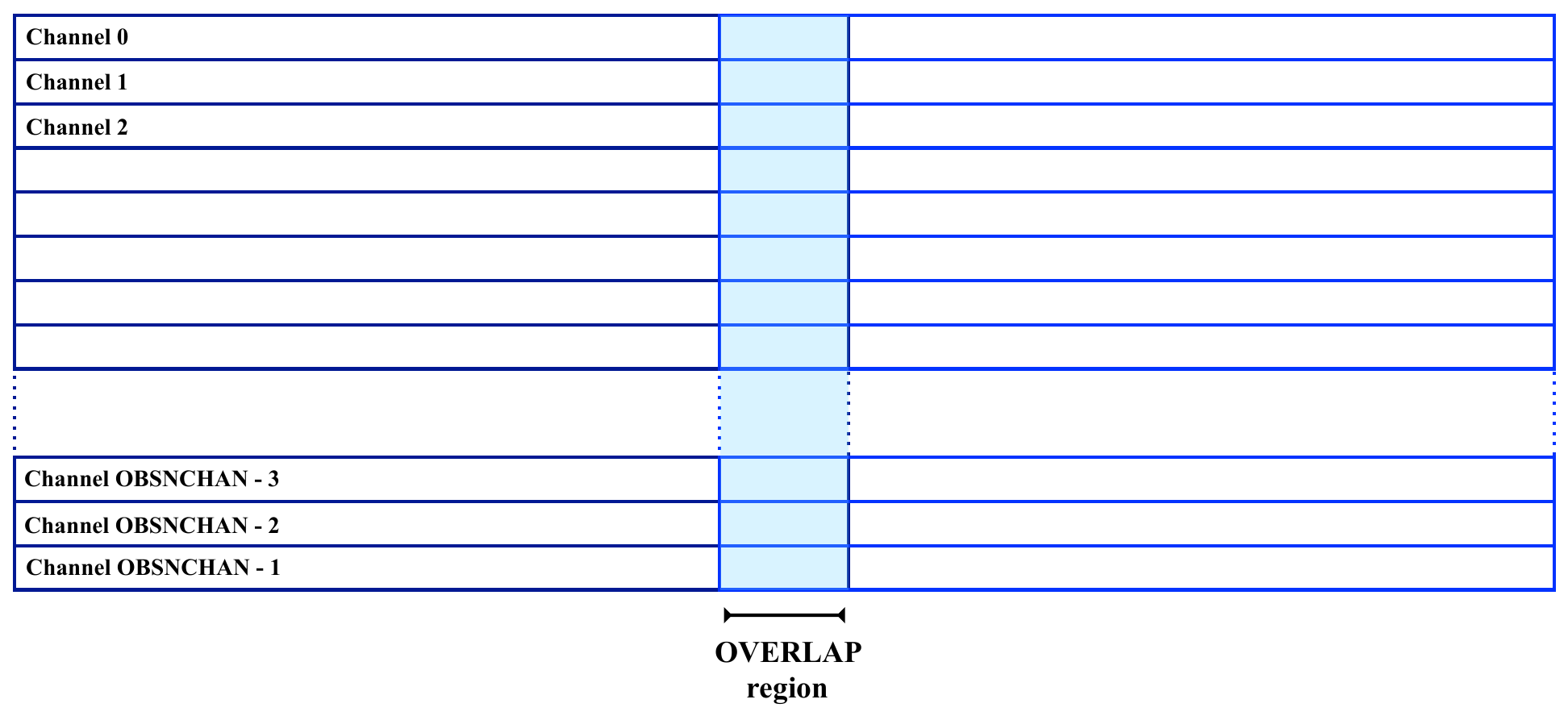}
\caption{\label{fig:overlap} The {\sc overlap} header field has units of samples.  Each sub-integration overlaps with the subsequent sub-integration by {\sc overlap} samples per channel. In general, this is 0 for all BL raw data.}
\end{figure*}

There are differences in block sizes in \bldr\ (keyword {\sc blocsize} in the file headers), due to
moving from older standards which accounted for overlapping blocks (Figure~\ref{fig:overlap}) to more optimal sizes, and in the case of Parkes Multibeam data, due to network bandwidth limitations (Table~\ref{tab:blocksize}). The differences in block size result in slightly different resolutions in the resulting filterbank data products, as detailed in \S~\ref{sec:filterbank} and Table \ref{tab:revchan}.
The raw file format is illustrated graphically in Figure~\ref{fig:rawchannels}.

The data portion of each block consists of voltages split into a number specified by the header keyword {\sc obsnchan}. Each coarse channel is $\sim 3$\,MHz wide, and made up of $\sim 512,000$ 8-bit real / 8-bit imaginary samples in two polarizations (32 bits, or 4 bytes, total per sample). The first pair of 8 bit values in the four byte set contains the real and imaginary values for channel 1, polarization 1, and time block 1. The second pair contains the real and imaginary values for channel 1, polarization 2, and time block 1. The subsequent bytes contain the same information for increasing time blocks. This pattern follows for the remaining channels (Figure~\ref{fig:rawchannels}).

\begin{deluxetable}{l l l l l l}
\tablewidth{0pt}
\tabletypesize{\scriptsize}
\tablecaption{\label{tab:blocksize}Data block sizes}
\tablehead {
\colhead{Revision} &
\colhead{Date ranges (or receiver)} &
\colhead{Integrations} & 
\colhead{Channels} &
\colhead{Bytes per sample} & 
\colhead{Total file size (bytes)}
}
\startdata
GBT Rev.~1A & 2016/1/1 - 2016/6/16; 2016/11/21 - 2016/11/22 & $512 \times 1009$  & 64 & 4 & 134,251,648\\
GBT Rev.~1B & 2017/2/10; 2017/3/29 & $512 \times 988$ & 64 & 4 & 129,499,136\\
GBT Rev.~2A & 2016/6/17 - present & $512 \times 1024$ & 64 & 4 & 134,217,728\\
Parkes & 1050CM & $512 \times 1024$ & 64 & 4 & 134,217,728\\
Parkes & Multibeam & $512 \times 1024$ & 44 & 4 & 92,274,688\\
\enddata
\end{deluxetable}

\subsection{Spectrograms\label{sec:filterbank}}

As noted above (\S~\ref{gbtparkes}), during the time between observing sessions, the raw data passes through the reduction and analysis pipeline, and in most cases the raw data are deleted to make space on the drives for the subsequent observing session. 

For the majority of targets, the raw data are passed to a 
GPU-accelerated code, \textsc{gpuspec}, to produce three files, each containing an array of integrated power as a function of frequency and time (Table~\ref{tab:dataproducts}). The input raw data are the same, but the output resolutions of the three files are optimized to support narrow-band SETI searches, pulsar observations, and astrophysical spectral line studies (although in practice they are not limited to these applications). 

The resultant files are written in Sigproc ``filterbank'' format, a relatively simple type of spectral data file with an origin in pulsar astronomy \citep{Lorimer:2006vv}. In order of operation, this code:
\begin{itemize}
  \item Reads data from file in GUPPI raw format.
  \item Performs three different fast Fourier Transforms (FFTs) with different
    numbers of channels.  This operation is performed on the GPU device, using
    the NVIDIA \textsc{cufft} library. 
  \item Squares the channelized output, add the two polarizations together, and integrates in time to form Stokes I data in GPU device memory.
  \item Writes output data to files in Sigproc filterbank format (Figure~\ref{fig:filterbank}).
\end{itemize}

\begin{figure*}
\centering
\includegraphics[width=0.8\textwidth]{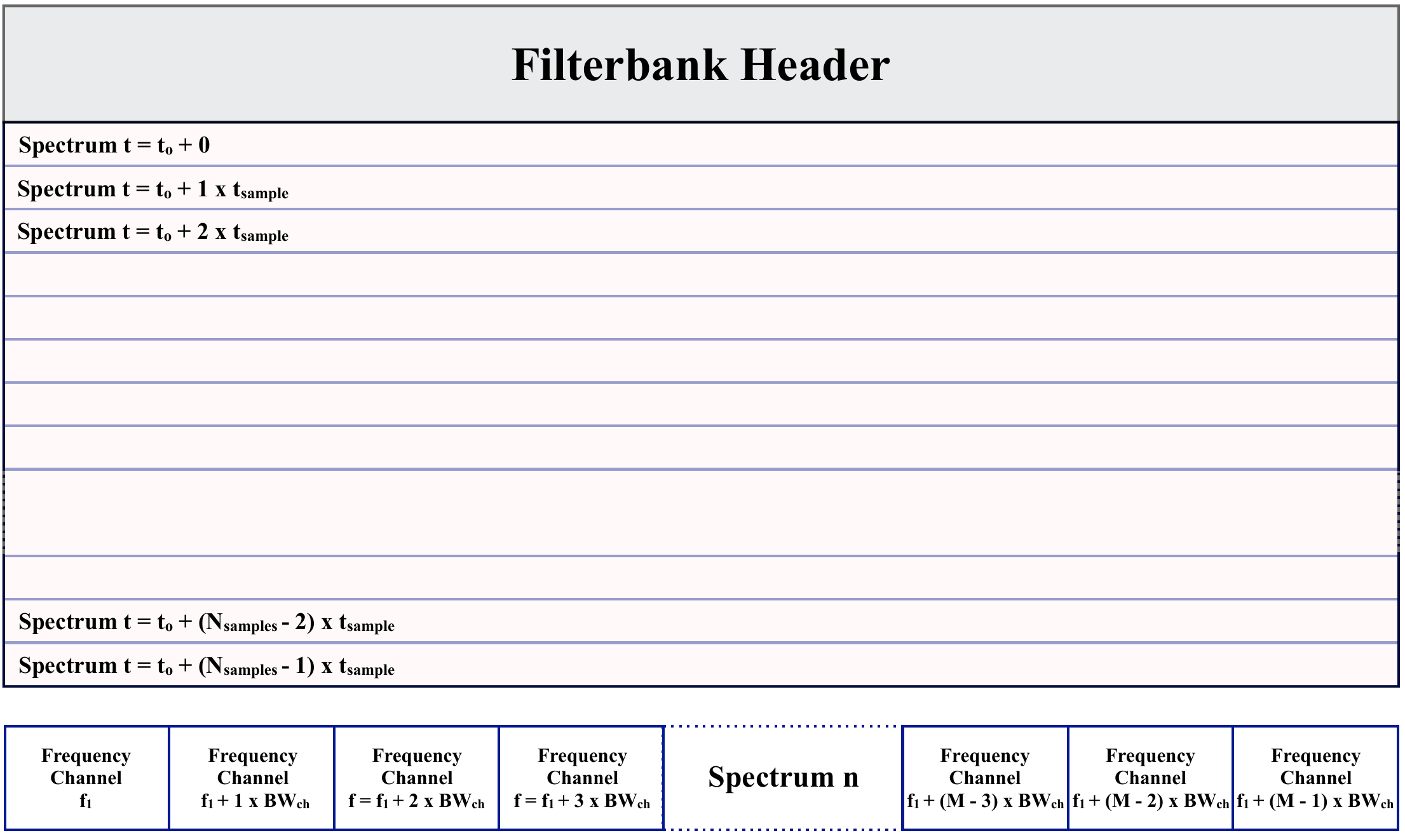}
\caption{\label{fig:filterbankformat}
Graphical representation of filterbank file, consisting of a series of spectra as a function of time.
}
\end{figure*}

The filterbank files are simply a header followed by a two dimensional array of integrated power (as 32-bit floats) as a function of radio frequency and time (as shown in Figure~\ref{fig:filterbankformat}). The filterbank format was chosen as it is well understood by already existing pulsar analysis tools. 

\begin{deluxetable}{l l l l l}
\tablewidth{0pt}
\tabletypesize{\scriptsize}
\tablecaption{\label{tab:dataproducts} Standard BL data products}
\tablehead {
\colhead{Product} &
\colhead{Purpose} &
\colhead{Filename Suffix} &
\colhead{Frequency Resolution} &
\colhead{Time Resolution}
}
\startdata
High Spectral Resolution (HSR) & SETI & .gpupsec.0000.fil & $\sim$3\,Hz & $\sim$18\,s \\
High Time Resolution (HTR) & Pulsar & .gpuspec.8.0001.fil & $\sim$366\,kHz & $\sim$349$\,\mu$s \\
Medium Resolution (MR) & Spectral line & .gpuspec.0002.fil & $\sim$3\,kHz & $\sim$1\,s \\
\enddata
\vspace{10pt}
Note: See Appendix~\ref{sec:filenaming} for more information about the filename suffixes.
\end{deluxetable}

\begin{deluxetable}{l c c l l}
\tablewidth{0pt}
\tabletypesize{\scriptsize}
\tablecaption{\label{tab:revchan} Channelization for GBT filterbank products}
\tablehead {
\colhead{Data Product} &
\colhead{FFT Length} &
\colhead{Integration Blocks} &
\colhead{Frequency Resolution} &
\colhead{Time Resolution}
}
\startdata
\multicolumn{5}{l}{Revision 1A (BLOCSIZE = 132251648)}\\
\hline
HSR & 1033216 & 51 & $\sim$2.84\,Hz  & $\sim$17.98\,s \\
HTR & 8 & 128 & $\sim$366.21\,kHz & $\sim$349.53\,$\mu$s \\
MR & 1024 & 3027 & $\sim$2.86\,kHz & $\sim$1.06\,s \\
\hline
\multicolumn{5}{l}{Revision 1B (BLOCSIZE = 129499136)}\\
\hline
HSR & 999424 & 51 & $\sim$2.93\,Hz  & $\sim$17.40\,s \\
HTR & 8 & 128 & $\sim$366.21\,kHz & $\sim$349.53\,$\mu$s \\
MR & 1024 & 2928 & $\sim$2.86\,kHz & $\sim$1.02\,s \\
\hline
\multicolumn{5}{l}{Revision 2A (BLOCSIZE = 134217728)}\\
\hline
HSR & 1048576 & 51 & $\sim$2.79\,Hz  & $\sim$18.25\,s \\
HTR & 8 & 128 & $\sim$366.21\,kHz & $\sim$349.53\,$\mu$s \\
MR & 1024 & 3072 & $\sim$2.86\,kHz & $\sim$1.07\,s \\
\enddata
\end{deluxetable}

The set of three filterbank products take up roughly 2\%\ the disk space of the original GUPPI raw data products. A small number of targets (e.g. ``special'' observations such as FRB\,121102, diagnostic targets, and pulsar observations) are denoted by adding the string `DIAG' to the target name in order to preserve the raw data products as discussed in Appendix~\ref{sec:filenaming}.

The data portion of any filterbank product is described by values in the file's metadata: \texttt{fch1},  \texttt{foff}, and \texttt{nchans}. These are the center frequency of the first fine channel of an integration, the frequency delta between adjacent fine channels, and the number of fine channels in any given integration. See Appendix~\ref{sec:fileheaders} for more details about filterbank metadata. 

For any given number of polyphase coarse channels, $\mathrm{N}_{\rm{c}}$, and number of fine channels, $\mathrm{N}_{\rm{f}}$, the center frequency of the first coarse channel ($\mathrm{F}_{\rm{cch1}}$) for any compute node is:

\begin{displaymath}
\mathrm{F}_{\rm{cch1}} = \mathrm{OBSFREQ} - \frac{\mathrm{N}_{\rm{c}}-1}{2\mathrm{N}_{\rm{c}}} \times \mathrm{OBSBW}
\end{displaymath}

where the GUPPI raw header values \texttt{OBSFREQ} and \texttt{OBSBW} are the center frequency of the raw file, and the width of the passband in the node, respectively.

The center frequency of the first fine channel, $\mathrm{F}_{\rm{fch1}}$, is given by:

\begin{displaymath}
\mathrm{F}_{\rm{fch1}} = \mathrm{F}_{\rm{cch1}} - \rm{floor}(\frac{\mathrm{N}_{\rm{f}}}{2}) \times \frac{\mathrm{OBSBW}}{\mathrm{N}_{\rm{c}}\mathrm{N}_{\rm{f}}}
\end{displaymath}


The resulting frequencies are shown in Table~\ref{tab:freqs}. For C-band and X-band, multiple banks of compute nodes are used for recording (e.g.\ for X-band, three banks numbered blc00 -- blc07, blc10 -- blc17, and blc20 -- blc27, although in practice any three of the eight available banks may be used). In order to preserve sensitive parts of the bandpass, frequencies overlap between adjacent banks.

Once reduced, the filterbank files are spliced (i.e.\ data from each of the 187.5-MHz-wide frequency bands from each compute node are combined) and collated onto local storage nodes at each respective observatory. 

Some existing analysis tools for pulsar/high time resolution searches prefer the filterbank files to have 8-bit integer values instead of the default 32-bit float. This conversion is done using the  \textsc{sum\_fil} program\footnote{\url{https://github.com/UCBerkeleySETI/bl_sigproc}}.

\subsection{Hierarchical Data Format - HDF5}

\label{sec:hdf5}
The Hierarchical Data Format version 5 (HDF5)\footnote{\url{https://support.hdfgroup.org/HDF5/}} is in wider use than filterbank (particularly outside of the scientific community). To improve the accessibility of BL data, as well as to take advantage of additional functionality provided by HDF5 (including improved data compressibility) we convert our filterbank files into HDF5 before making them publicly available.

We have developed a software suite\footnote{\url{https://github.com/UCBerkeleySETI/blimpy}}, {\sc blimpy} (Breakthrough Listen I/O Methods for PYthon), to import and display filterbank, HDF5, and GUPPI RAW data in Python \citep{ascl:blimpy}. {\sc blimpy} also contains tools for conversion of filterbank to HDF5. {\sc blimpy} will be described in more detail in an upcoming paper.

As part of the conversion from filterbank to HDF5 we check pointings against historic telescope records and adjust fine channel offsets. Originally reduced data products were created with first channels incorrectly labeled at their center frequencies instead of at the edge; these half-channel errors only affect the headers and not the data. We also trim away the edges of the frequency range where the bandpass sensitivity falls off, thus significantly reducing the size of the data using the {\sc blimpy} \texttt{bldice} command. 

In general, we trim the GBT L-band data down to the $1025 - 1925$\,MHz range, GBT S-band data to $1820 - 2720$\,MHz, and Parkes 10CM to $2574 - 3444$\,MHz. Some files contain smaller ranges of frequencies; reasons for this may include:

\begin{itemize}
  \item Early on in the project files were trimmed by including only data from the central four compute nodes recording the most sensitive parts of the bandpass
  \item Due to intermittent system/disks failures while building out each backend system, data from some nodes were unusable, and thus certain ranges are missing.
  \item Occasional tuning errors caused the center frequency to be off by enough to bring the bandpass outside of the recorded range.
\end{itemize}

\subsection{Instrumental Effects and Data Artifacts}\label{sec:artifacts}

At both GBT and Parkes, the complete BL signal chain includes numerous components, from radio receivers, through analog signal downconversion, digital sampling, channelization, retransmission, recording and finally software post processing.  Virtually all of these components impose instrumental effects on the resultant data products, and many of them suffer sporadic faults.  These issues are compounded by instrument commissioning, testing and continuous software development.  Here we detail several of the most prominent instrumental features found in \bldr\ and highlight known data artifacts affecting a small percentage of the collection.  We note that this list is not exhaustive, and caution should be taken in ruling out previously undiscovered instrumental problems when encountering anomalies in the data set.

\subsubsection{Radio Frequency Interference}

Radio telescopes are highly susceptible to RFI, from local transmitters as well as both airborne and spaceborne sources. Such interference can dominate the data, given its power, perseverance, and multitude of forms.  Often the strength of these transmitters is sufficient to cause their apparent power to spread well beyond the time and frequency extent of the specific transmission due to analog saturation and digital overflow. It is difficult to overstate the pervasiveness, diversity and challenge of RFI in searches for radio technosignatures.

Data collected from the single pixel receivers use a strategy depicted in Section~\ref{sec:observingstrategy} to help filter out local interference. Further information about RFI detection and mitigation of spectrograms included in this data release is presented in \citePrice{}. 

\subsubsection{Analog and Digital Filtering}

\begin{figure*}
\centering
\includegraphics[width=0.32\linewidth]{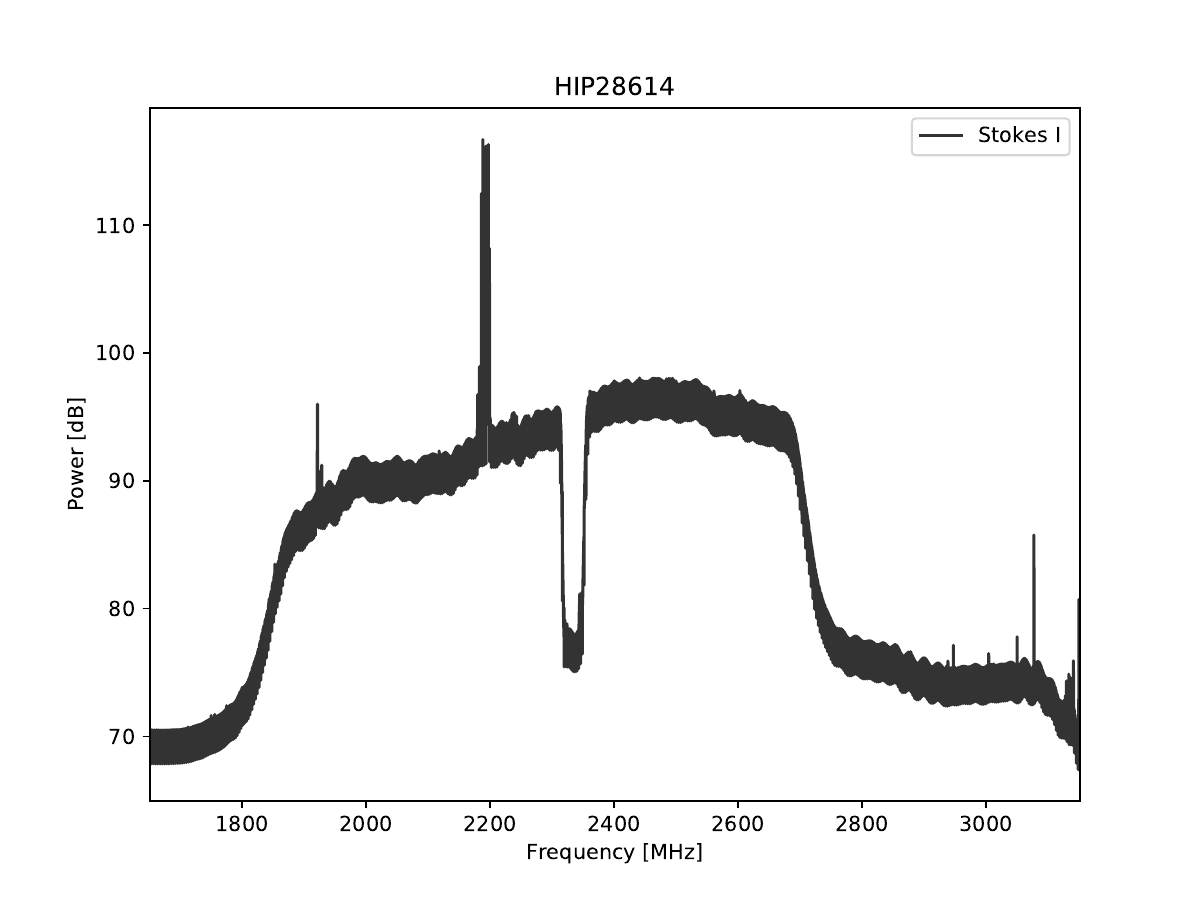}%
\includegraphics[width=0.32\linewidth]{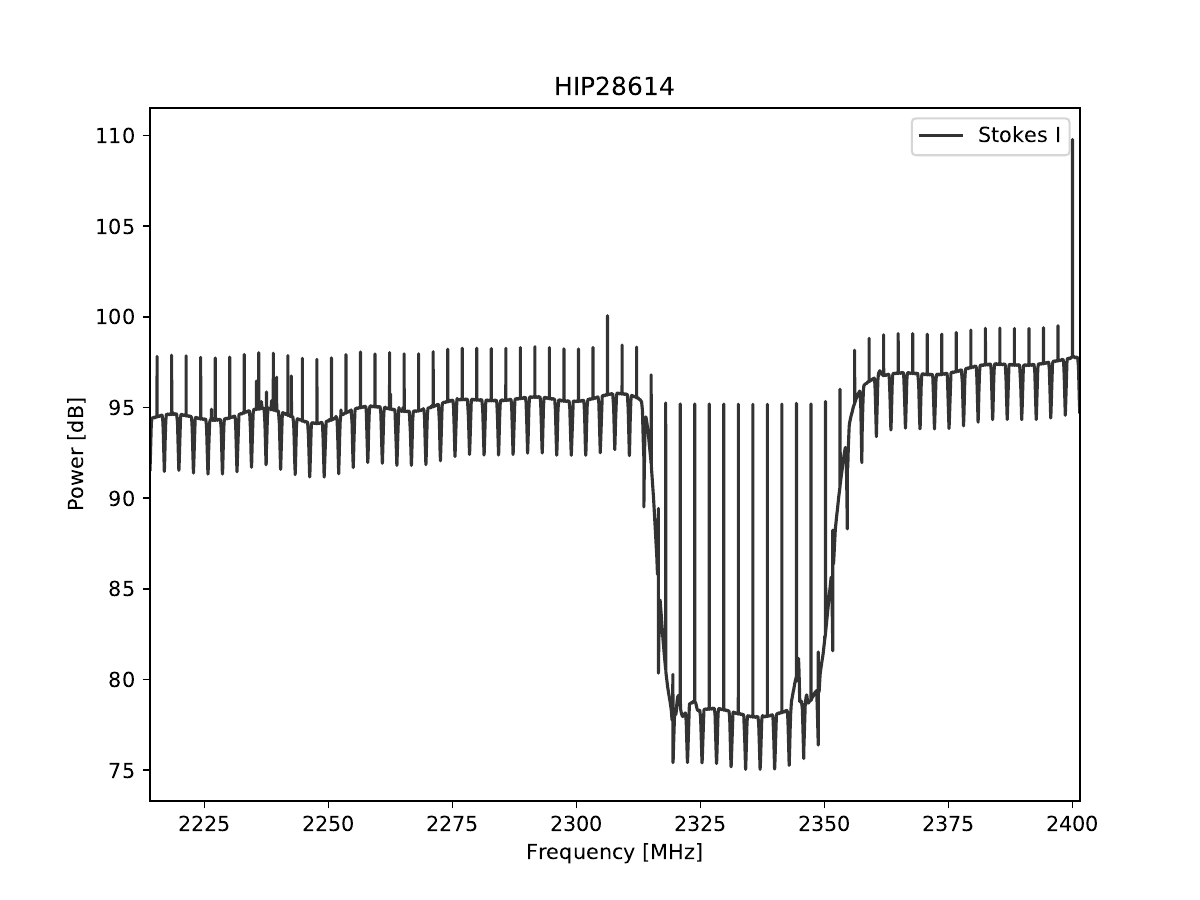}
\includegraphics[width=0.32\linewidth]{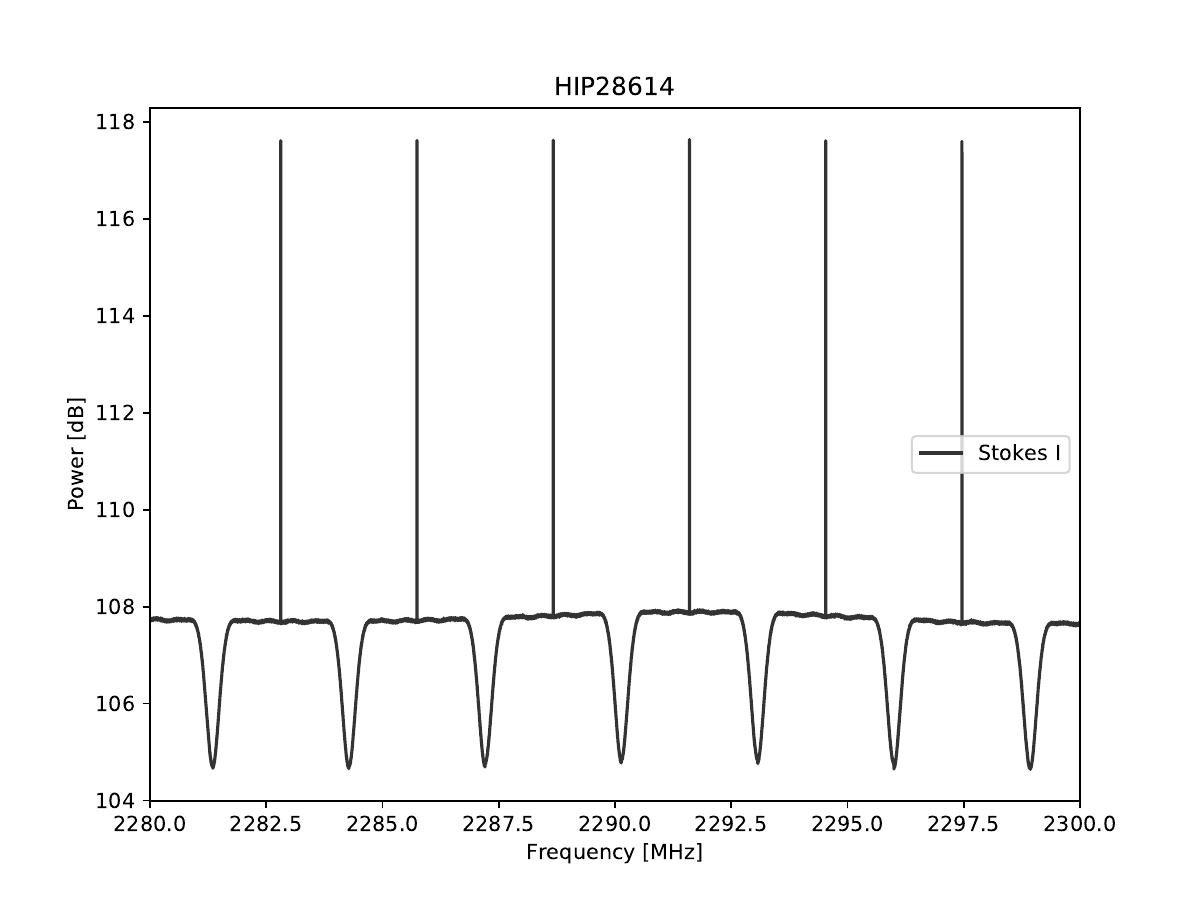}
\caption{\label{fig:filspec} One-dimensional S-band spectra (collapsing the time dimension from the spectrogram) plotted using the same data for HIP\,28614 from the top left panel of Figure~\ref{fig:filterbank}. The left panel shows the spliced data for all 8 compute nodes. The dip around $\sim 2330$\,MHz is due to the observatory's S-band notch filter (which is used to reduce interference from Sirius and XM satellite transmissions). Bright RFI is also visible at $\sim 2200$\,MHz. The middle panel zooms in to show 1/8 of the data (the 187.5\,MHz from compute node blc04), showing the lower-gain region corresponding to the notch filter in more detail. The 64 individual 3-MHz-wide coarse channels are visible, each with a fall-off at each edge of the coarse channel, and each with a spike in the DC bin at the center of the coarse channel. Each of the coarse channels is more finely channelized by the {\sc gpuspec} FFT: 8 fine channels per coarse channel for the HTR data product, 1024 fine channels per coarse channel for the MR (the data product plotted here), and $1,048,576$ fine channels per coarse channel for the HSR data product (Table~\ref{tab:dataproducts}). The right panel is zoomed into the 2280--3000 MHz range to better depict the polyphase shape of roughly 7 coarse channels. }
\end{figure*}

The incoming analog bandpass is not flat and different for all receivers. These are described in the respective Green Bank\footnote{GBT Observer's Guide - \url{https://science.nrao.edu/facilities/gbt/observing/GBTog.pdf}} and Parkes observing guides\footnote{Parkes Radio Telescope Users Guide - \url{http://www.parkes.atnf.csiro.au/observing/documentation/user_guide/pks_ug.pdf}}. The BL instrument also imposes its own polyphase filter shape on the data \citep{2016arXiv160703579P}. Each coarse channel ($\sim 3$\,MHz) has a reduction in sensitivity toward the channel edges, and a DC channel in the center. A typical coarse channel shape (along with some RFI) is shown in Figure~\ref{fig:filspec}. There are no corrections for these in the public data. 

\begin{figure*}
\centering
\includegraphics[width=0.32\linewidth]{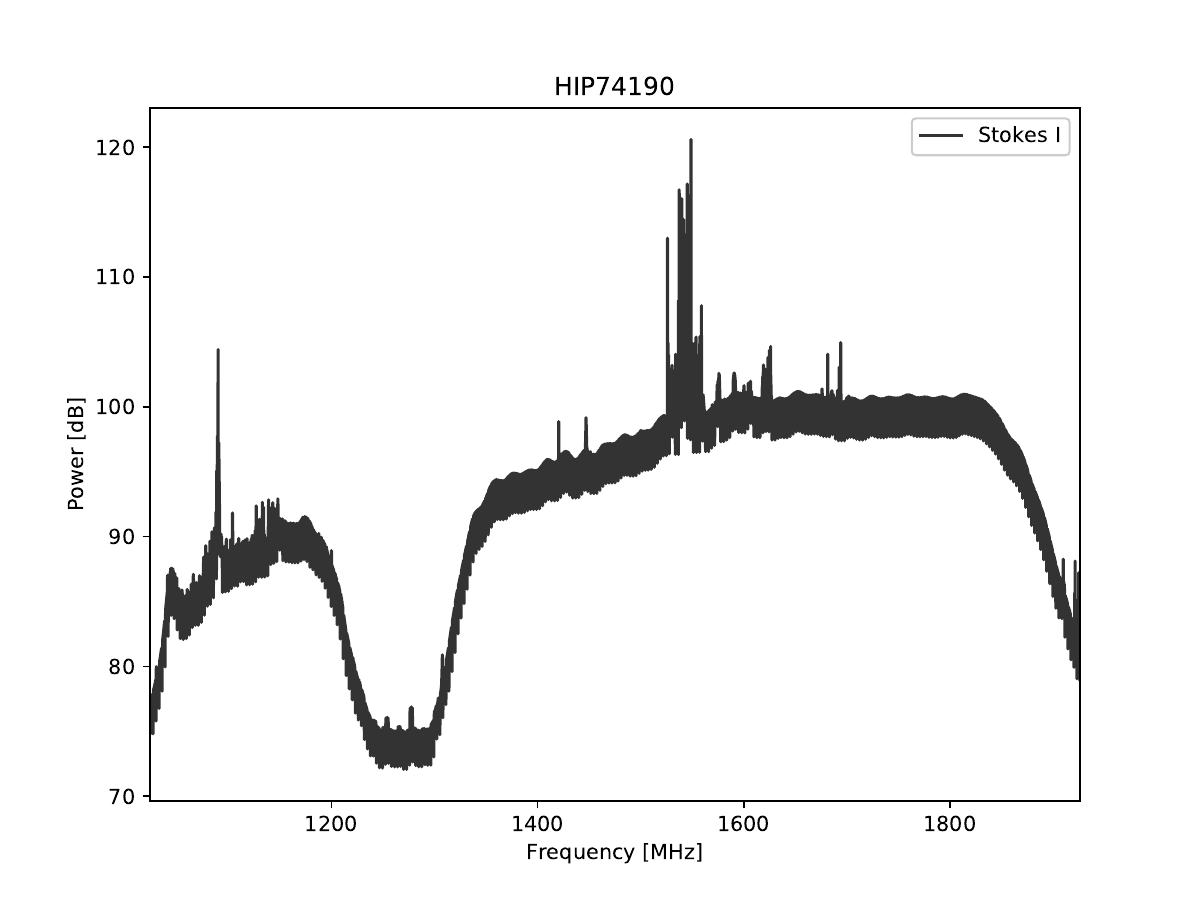}%
\includegraphics[width=0.32\linewidth]{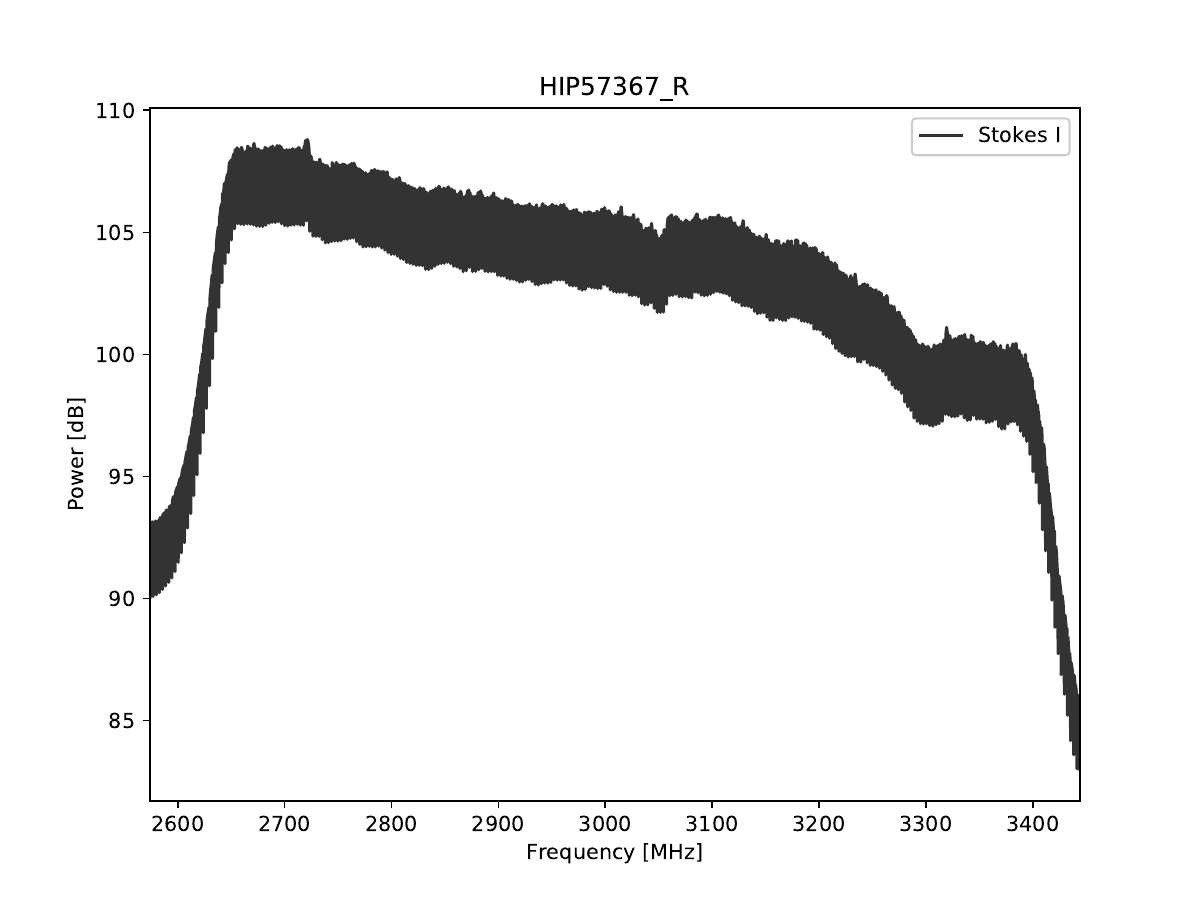}
\caption{\label{fig:morefilspec}Examples of spectra similar to Figure~\ref{fig:filspec} for both the L-band receiver at Green Bank (left) and the Parkes 10CM receiver (right). The L-band receiver has a notch filter between 1.2--1.34\,GHz to reduce interference from nearby air surveillance radar. These examples are also trimmed to a smaller range of frequencies, as described in Section~\ref{sec:hdf5}.}
\end{figure*}

\subsubsection{Dropped Blocks and Lost Data}

\begin{figure*}
\centering
\includegraphics[width=0.5\linewidth]{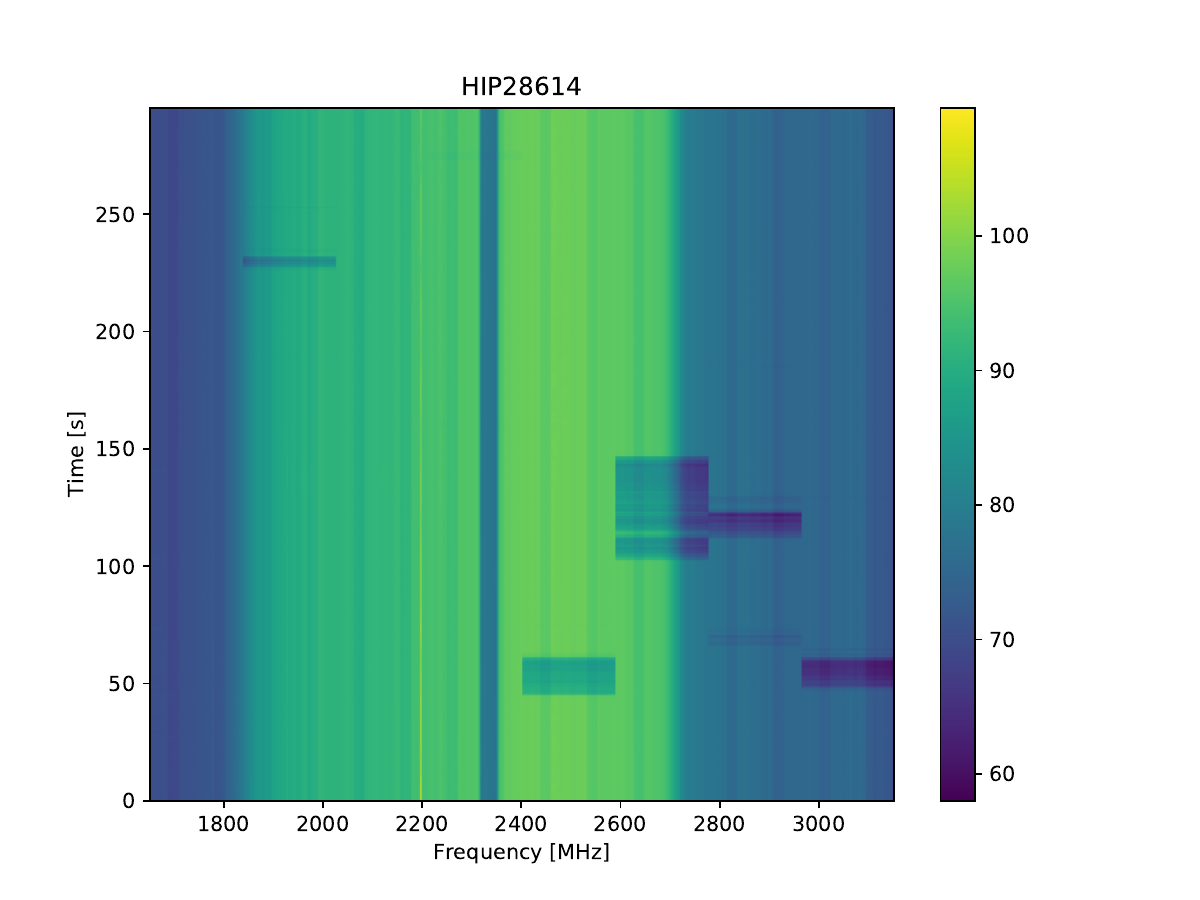}%
\includegraphics[width=0.5\linewidth]{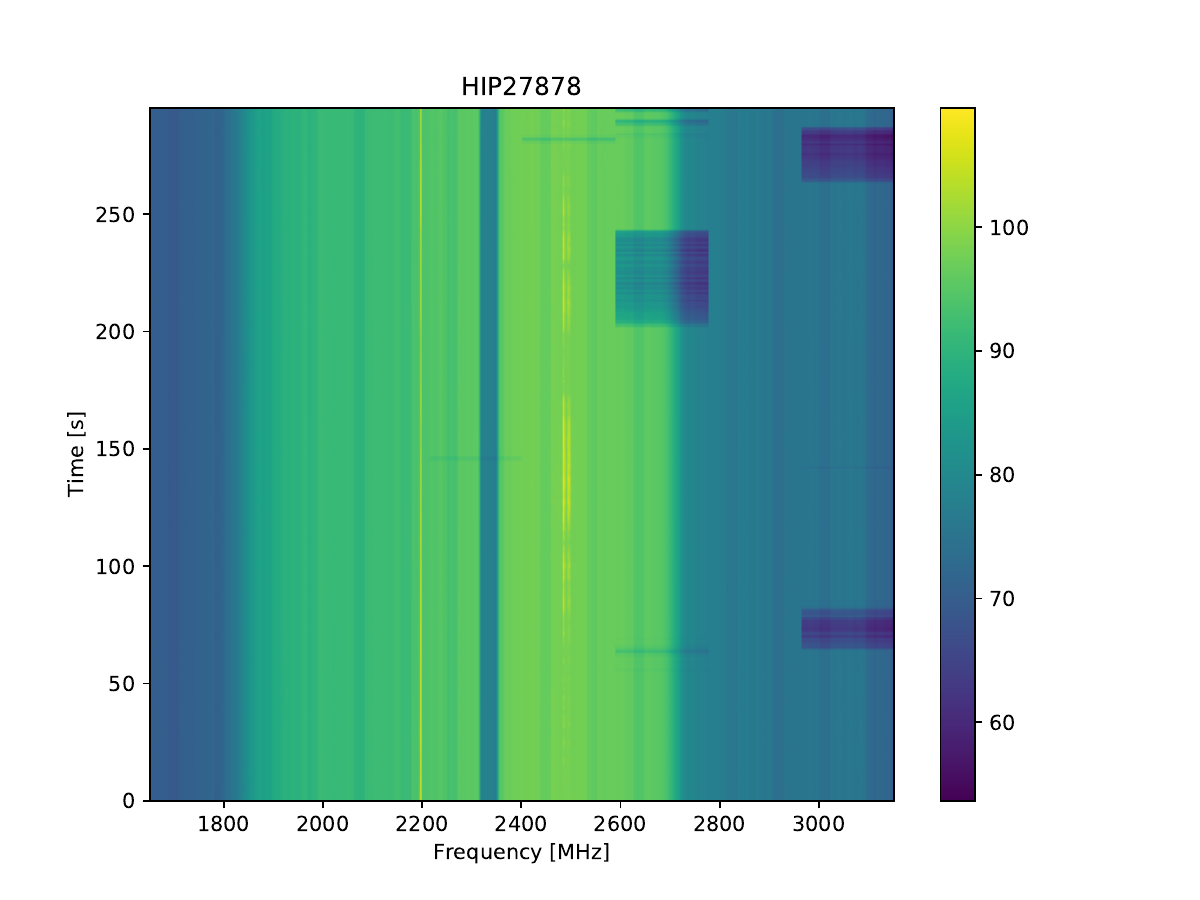}
\caption{\label{fig:filterbank2}These are targets from one of a small number of observational cadences affected by patches of reduced sensitivity due to dropped ethernet packets during the course of an observation. The affected data are visible as blue rectangles spanning 1/8 of the total band of a given 1500 MHz band of eight compute nodes, i.e.\ the 187.5\,MHz associated with individual nodes. The length in time (or ``height'') of these rectangles is arbitrary, as it only happens when a particular node is missing data due to several factors that are hard to predict or control. Most datasets in \bldr\ are not affected in this manner, but these dropped packets are not to be confused with expected reduced sensitivity due to notch filters (such as the one shown here from 2300-2360\,MHz) and the areas outside of the bandpass (i.e. below 1800\,MHz and above 2700\,MHz in this case).}
\end{figure*}

Occasionally other issues affecting data quality arise. During observations each compute node receives about 1 GB\,s$^{-1}$ of data from the telescope. Hardware issues (failing hard drives, degraded RAID arrays, faulty switches, bad cables, xfs partition fragmentation which reduces write speeds even when using Direct I/O) can impede the ability of the compute nodes to record data at this rate, resulting in dropped blocks (short spots of missing data). This causes an occasional reduction in sensitivity, the effects of which are shown in Figure~\ref{fig:filterbank2}.

More rarely, problems with the communication between the BL backend and the telescope may result in gaps of 187.5\,MHz where data are completely lost for a particular compute node.

In addition to these occasional hardware or software issues, bad weather may reduce sensitivity, or cause observations to stop mid-scan to park the telescope for safety. Additionally, observations at the end of each BL session may consist of less than five minutes of data. 

\subsubsection{Two-bit requantization of raw data}

During early commissioning at GBT (approximately MJD 57402--57502) we requantized the raw data from 8-bit to 2-bit (1 bit real / 1 bit imaginary) using the \textsc{raw\_quant} program\footnote{\url{https://github.com/UCBerkeleySETI/gbt_seti}} in order to reduce disk usage. The methodology of this process is described below. This step is no longer performed except when creating data to be copied offsite for the SETI@home project (Figure~\ref{fig:pipeline})). Data requantized to 2-bits shows a significantly diminished analog bandpass response, has reduced sensitivity and issues with saturation in the presence of strong sources. 

Four-level quantization taken from \cite{thompson1986interferometry} Table A8.1 – divides a Gaussian distribution into 4 bins:

\begin{center}
\begin{tabular}{rcl}
    -N & = & sample $<$ -vo  \\
    -1 & = & 0 $>$ sample $>$ -vo \\
     1 & = & 0 $<$ sample $<$ vo \\
     N & = & sample $>$ vo
\end{tabular}
\end{center}

Where -N, -1, 1, N are the voltage values represented by each quantized sample and vo is the unquantized threshold.

From the table, optimal efficiency for this scheme is achieved by taking vo = 0.98159883 * rms and N = 3.3358750. Theoretical efficiency in this case is given as $\sim$0.8825.

We apply this scheme in the following way:

\begin{itemize}
\item{Open a raw guppi data file and calculate the rms and mean for each channel-polarization in the first sub integration (~1 sec)}
\item{Subtract off the mean from each sample, and bin according to the thresholds above.}
\end{itemize}

We'll assign a 2 bit value to each of the above bins as follows:

\begin{center}
\begin{tabular}{ccc}
		Binary (LSB first) & Unsigned Int &	Quantized Value\\
		\hline
		00		&	0		&	N \\
		10		&	1		&	1 \\
		01		&	2		&	-1 \\
		11		&	3		&	-N \\
\end{tabular}
\end{center}

\section{Optical: BL on APF}\label{apf}

The Automated Planet Finder Telescope \citep[APF;][]{Radovan2014} is a 2.4-meter optical telescope at Lick Observatory in California. Its primary aim is to search for earth-like extrasolar planets. The single instrument is a $R \sim 95,000$ optical echelle spectrometer with wavelength coverage from 374 to 950\,nm. BL uses the APF for 36 nights per year and targets nearby stars with a maximum exposure time of 20 minutes (aiming for an optimal signal-to-noise ratio of 100:1). Three spectra are obtained for each star in order to more easily distinguish cosmic ray hits from interesting signals. The nominal decker projects to a sky angle of $1.0\arcsec\ \times 3.0\arcsec$, which at 10\,pc corresponds to $10 \times 30$\,AU, allowing the detection of laser emission from planets that are offset from their stars, as well as emitters that may lie in the foreground or background within the line of sight \citep{Tellis2015}. 

\begin{figure}
\centering
\includegraphics[width=0.5\linewidth]{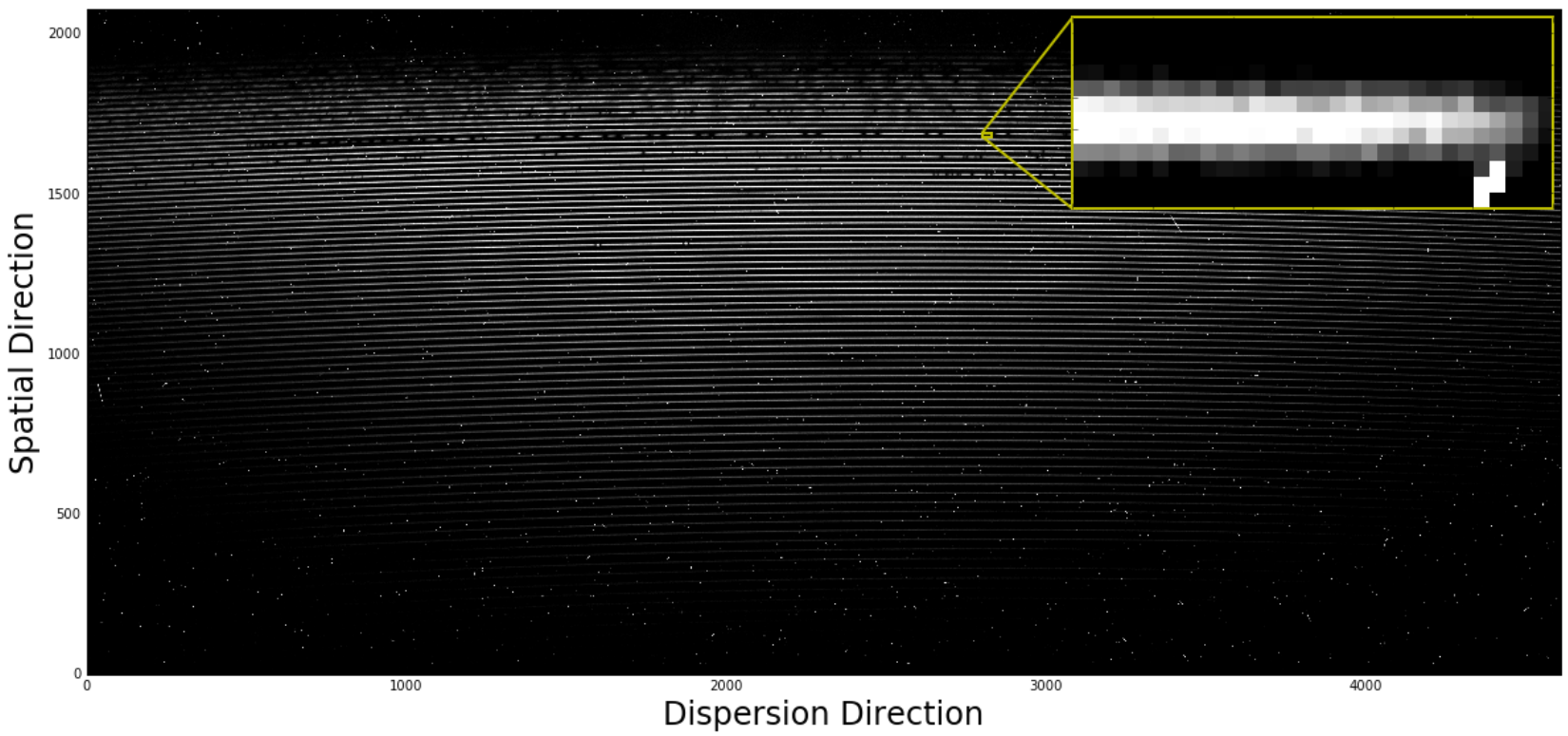}
\caption{A raw 2D echelle spectrum (figure from \citealt{lipman:19}) of Boyajian's Star, 4,608 pixels in the dispersion direction and 2,080 pixels in the spatial direction, containing 79 spectral orders. The zoomed image in the upper-right corner is a segment of one of these spectral orders, which shows a cosmic ray in the bottom right corner. The wavelength coverage of the Levy is $373.0 - 1020.6$\,nm; however CCD fringing limits the usability of the reddest orders.}
\label{fig:tabby2d}
\end{figure}

\begin{figure}
\centering
\includegraphics[width=0.9\linewidth]{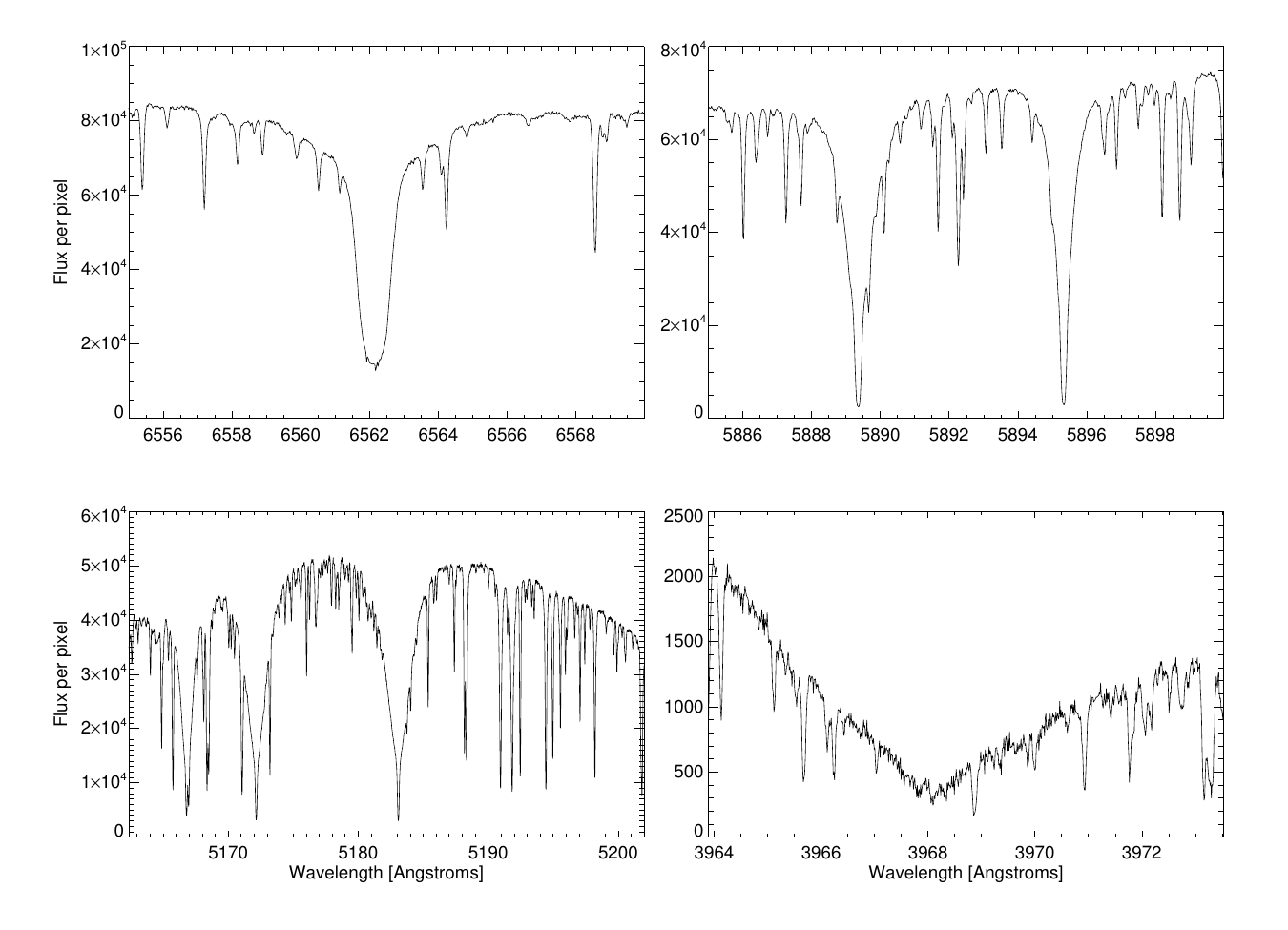}
\caption{Four prominent absorption features extracted from a reduced, 1D, high-resolution spectrum of HD\,10700 from the APF. The resolution of 95,000 allows for identification of potential laser lines emitted along the line of sight towards the target star. Clockwise from top left: H-alpha, the Sodium-D doublet, the Magnesium B triplet and the Calcium II H-line showing broad absorption. Fluxes are in photons per pixel.}
\label{fig:apf_1d}
\end{figure}

The two standard data products available from the APF are the raw 2D image (each 18.3\,MB), an example of which is shown in Figure~\ref{fig:tabby2d}, and the extracted 1D spectrum (each 1.4\,MB), as shown in Figure \ref{fig:apf_1d}. These files adhere to the common FITS standard (\citealt{fits,fits3}) and are thus easily readable by existing tools. 

BL utilizes both the raw and reduced images to search for monochromatic laser lines that are narrow in wavelength space and match the point-spread-function (in the raw images) and line-spread-function (in the reduced images) of the instrument. Typical searches by BL are described by \citet{Tellis2017} and \citet{lipman:19}, including details of cosmic ray removal, telluric absorption and emission lines, and blaze function correction. Each BL night is run in collaboration with the California Planet Search and utilizes scheduling software described by \citet{Holden2016}.

As with the radio searches of nearby stars, targets are selected from the sample of \citet{Isaacson2017}. Files for a total of 789 targets are available as part of \bldr. A tutorial, a description of headers and data formats, as well as example files are also available online\footnote{\url{https://github.com/UCBerkeleySETI/breakthrough/tree/master/APF}}. 

\section{The Breakthrough Listen Public Data Archive}\label{archive}

\begin{figure}
\centering
\includegraphics[width=0.5\linewidth]{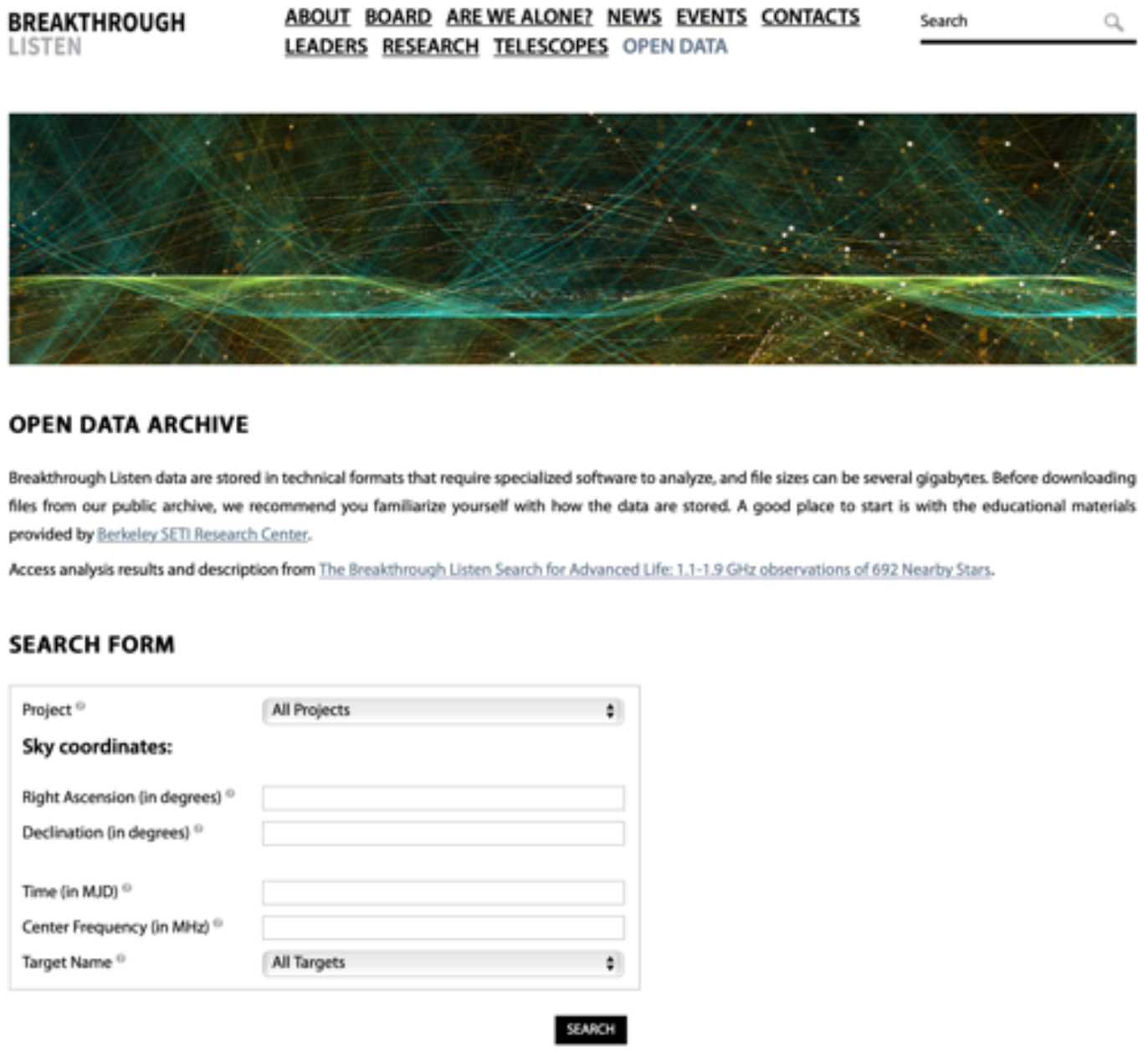}\hspace{0.05\linewidth}%
\includegraphics[width=0.4\linewidth]{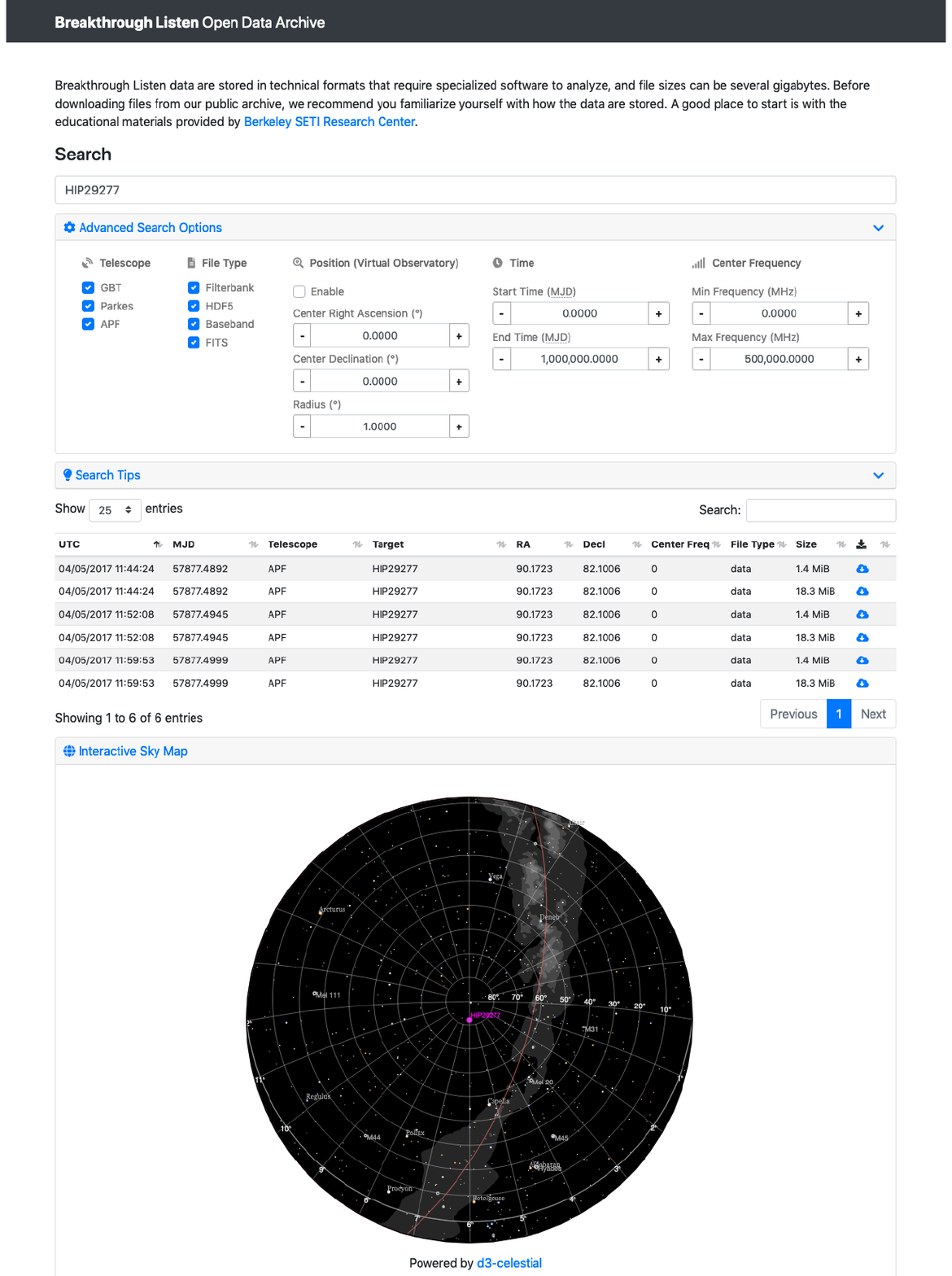}
\caption{Left: Open Data Archive hosted on the Breakthrough Initiatives web site. 
Right: Beta interface to the same dataset, with additional search and visualization options, as hosted by Berkeley SETI Research Center.
}
\label{fig:opendataweb}
\end{figure}

The BL data archive strategy is designed to facilitate analysis by the internal science team as well as engagement by external parties. Key information about the public data products is kept in a MySQL database containing basic metadata and URLs for all available files. Frontends to this database are provided via a portal on the Breakthrough Initiatives website\footnote{
\url{https://breakthroughinitiatives.org/opendatasearch}
}, and also by a beta interface with additional functionality hosted by Berkeley SETI Research Center\footnote{\url{http://seti.berkeley.edu/opendata}} (Figure~\ref{fig:opendataweb}). Data can be searched by target name, times, frequencies, or location.

The majority of data in \bldr\ are stored on 11 commodity public data servers at UC Berkeley, each with roughly 200\,TB of usable storage and its own 10\,Gb\,s$^{-1}$ link to the internet. Having a fleet of separate servers increases parallelization as well as uptime. Since the database can link to files hosted at any publicly-accessible URL, we can also seamlessly link to data at national computing centers and commercial cloud providers. We anticipate our use of cloud storage for BL data may increase in future.

Additional data, not yet available in \bldr, are still located on storage hardware at the telescope sites. Network bandwidth limitations (a few Gb\,s$^{-1}$) at the telescope sites prevent us from incorporating files stored there into the archive directly, but as they are transferred to public servers on fast connections at UC Berkeley (or in the cloud), the BL public data archive will continue to grow.

In future we also anticipate making increasing amounts of data available in other popular community-supported formats such as SigMF \citep{West_19:fosdem}, or as images appropriate for machine learning algorithms. 

The total volume of the archive is expected to grow to approximately 25\,PB over the next several years, and will ultimately include data from many facilities. 

We are also exploring technologies such as Ceph\footnote{\url{https://ceph.com}} and Globus\footnote{\url{https://globus.org}}, which in combination with more flexible downloads enabled by HDF5 will allow users to download portions of files rather than necessitating an entire file (which can be up to $\sim 100$\,GB in size) to be retrieved from the archive.

\section{Summary}
The large data volumes, complex instrumentation and the intricacies of varied file formats make acquisition, processing, and storage of BL data a challenging undertaking. With the release of \bldr\ ,the BL open data archive currently contains almost 1\,PB of data.  These data include the complete set of 1.10--3.45\,GHz observations used in the technosignature search presented by \citePrice{} as well as raw and reduced data from a number of particularly interesting targets for both SETI and general astrophysics. This is further quantified and described in Appendix~\ref{app:summary}. Data continue to be acquired at APF, GBT and Parkes, including new observing modes and radio receivers. Future BL Data Releases will include these data, as well as data collected from new BL facilities such as MeerKAT. With \bldr\ and future releases, it is our hope that experts from outside the SETI community, and even from outside radio astronomy, will be empowered to help us improve our technosignature searches and increase the general scientific return from BL observations. The process of making these data available has already revealed to us avenues for improvement in both organization and searchability - we will continue to strive for advancing the general availability and usability of our data. 

\section*{Acknowledgements}

Breakthrough Listen is managed by the Breakthrough Initiatives, sponsored by the Breakthrough Prize Foundation.
The Parkes radio telescope is part of the Australia Telescope National Facility which is funded by the Australian Government for operation as a National Facility managed by CSIRO. The Green Bank Observatory is a facility of the National Science Foundation, operated under cooperative agreement by Associated Universities, Inc. 
We thank the staff at Parkes and Green Bank observatories for their operational support.
We are very grateful for the donations of the Levy family that helped facilitate the construction of the Levy spectrograph on APF. Research at Lick Observatory is partially supported by a generous gift from Google.

\facilities{APF, GBT, Parkes}



\appendix

\section{File Naming Conventions}\label{sec:filenaming}

\subsubsection{Raw}
For GBT:\\
\texttt{\bb{blc\bc{<Node \#>}\_guppi\_\bc{<MJD Day>}\_\bc{<MJD Seconds>}\_\bc{<Target>}\_\bc{<Scan \#>}.\bc{<File \#>}.raw}}

For Parkes:\\
\texttt{\bb{blc\bc{<Node \#>}\_guppi\_\bc{<MJD Day>}\_\bc{<MJD Seconds>}\_\bc{<Packet Index>}\_\bc{<Target>}\_\bc{<Scan \#>}.\bc{<File \#>}.raw}}
\\

\textbf{Node \#}: As data are being recording, the filenames begin with the `guppi\_' prefix. Once reduced into filterbank files, we add this `blc' prefix (along with the number of the pertinent compute node) in order to denote successful completion of the reduction, as well as remove file name collision with similarly named files on other compute nodes.

\textbf{MJD Day, MJD Seconds, and Packet Index}: Next is `MJD Day' which is the integer portion of the Modified Julian Date at the beginning of this scan, followed by `MJD Seconds' which is the number of seconds of midnight UTC. Filenames of early BL GBT data products do not contain `MJD seconds'. Due to naming collisions of multiple observations of the same star in the same day, this was added at the end of 2016 January. At Parkes there is an additional integer field inserted in the filename between `MJD seconds' and `Target' relating to current packet index, that is used for debugging the BL backend.

\textbf{Target}: The target name may contain underscores (making parsing difficult). Diagnostic sources (e.g.\ pulsars) also contain the string `DIAG'; the processing pipeline recognizes this tag and preserves raw data which would otherwise be deleted after filterbank files have been generated. These raw data are kept for continued analysis, since they contain the full resolution, as well as amplitude and phase information, that are lost on conversion to spectrograms. In all other cases the raw data are immediately deleted after the filterbank files have been generated.

\textbf{Scan number}: These are provided by the observatory, start at 0001 at the beginning of each session, and increment for each scan; due to diagnostics and interruptions numbers are not always sequential in the BL public datasets.  

\textbf{File number}: Each raw file corresponds to $\sim 22.9$\,s of observation (128 blocks per file at 0.179\,s per block), so a typical 5-minute observation is split into 14 raw files, numbered sequentially. The file number starts at 0000 and increments by 1 for each 128-block file. A 128-block file is about 16\,GB. A partial raw file at the end of the series will contain enough full blocks to complete the exact 5-minute scan and so is typically smaller than 16\,GB. 

\subsubsection{Filterbank}

Same as raw above, except \texttt{.raw} is replaced by: \texttt{\bb{.gpuspec.\bc{<Resolution Type>}.fil}} where \texttt{<Resolution Type>} (noted in Table~\ref{tab:dataproducts}) is one of the following:
\begin{itemize}
\item \texttt{0000} : high spectral resolution (HSR) 
\item \texttt{0001} : high time resolution (HTR) 
\item \texttt{8.0001} : high time resolution (HTR), but 8-bit integers (a data type required, or preferred, by some analysis tools) instead of the default 32-bit floats 
\item \texttt{0002} : medium spectral/time resolution (MR)
\end{itemize}

Some data products in \bldr\ were processed with an updated version of {\sc gpuspec} called {\sc rawspec} which has been optimized for speed and better use of GPU memory. In such cases ``gpuspec'' in the filenames is replaced by ``rawspec''. 

The filterbank products from the separate compute nodes are typically spliced together into a single file. The names of these single files contain the word ``spliced", along with an ordered list of the processing nodes which went into making up these collated files. In a small fraction of cases where nodes experienced problems during recording for a particular target, lost data from the missing nodes are padded with zeros, and the node numbers in the filename replaced with `zz'. Unused overlapping nodes are also removed, and denoted `oo' in the output filenames.

\subsubsection{HDF5}

\texttt{\bb{\bc{<Telescope>}\_\bc{<MJD Day>}\_\bc{<MJD Seconds>}\_\bc{<Target>}\_\bc{<Resolution>}.h5}}
\\

Since the HDF5 version is the ``final" version, we remove some diagnostic information from the filenames, renaming them for simplicity.

`Telescope' identifies the telescope where the data were taken (GBT, Parkes, or future telescopes, e.g. MKat for MeerKat). `MJD,' `MJDsec,' and `Target' are the same as before, and as such the target name may include underscores. `Resolution' is either ``fine" for fine spectral resolution, ``time" for fine time resolution, and ``mid" for mid spectral/time resolution. Names are globally unique, human readable, and computer parsable.

\subsubsection{FITS (APF)}

2D echelle spectrogram:\\
\texttt{\bb{ucb-\bc{<Session Index>}\bc{<Scan \#>}.fits}}

1D extracted spectrum:\\
\texttt{\bb{r\bc{<Session Index>}.\bc{<Scan \#>}.fits}}


\section{Headers and Metadata}\label{sec:fileheaders}

Below we describe the metadata describing the payload within any of our data products. For the raw voltage files there are ASCII headers at the beginning of every block described below. For the reduced filterbank or HDF5 files the non-ASCII metadata also described below are viewable with various utilities, most easily with the {\sc blimpy} \texttt{watutil} command. 

\subsection{Raw Headers}

\begin{itemize}
\item [] \texttt{ACC\_LEN}: Accumulation length (\# of spectra averaged together in time) - set to 1 for coherent modes only.
\item [] \texttt{AZ}: Azimuth of the telescope
\item [] \texttt{BACKEND}: A name for the backend (or backend compatibility) of the recorded data.  Breakthrough Listen canonically uses \texttt{GUPPI} for maximizing backwards compatibility with legacy code.
\item [] \texttt{BANDNUM}: For telescopes with multiple IF downconverters (aka ``bands"), this field identifies which IF band provided the data for the given RAW file.
\item [] \texttt{BANKNAM}: Name of the Breakthrough Listen bank where this file was recorded, labeled by the ``Player'' name (see \citealt{gbtinstrument})
\item [] \texttt{BANKNUM}: Once channelized, different parts of the IF band are received and processed by different backend hosts.  This field identifies which portion of the IF band is present in a given RAW file.
\item [] \texttt{BASE\_BW}: Occupied portion of the Nyquist bandwidth (no longer used).
\item [] \texttt{BINDHOST}: Ethernet device for receiving data
\item [] \texttt{BLOCSIZE}: The size of the following raw data segment, in bytes. Does not include padding bytes from Direct I/O. Can be expressed as $2 \times \mathrm{NPOL} \times \mathrm{NTIME} \times \mathrm{NCHAN} \times \mathrm{NBITS} / 8$
\item [] \texttt{BMAJ}: Full width half maximum of the major axis of the beam
\item [] \texttt{BMIN}: Full width half maximum of the minor axis of the beam
\item [] \texttt{CAL\_DCYC}: Duty cycle for pulsed calibrator
\item [] \texttt{CAL\_FREQ}: Pulse frequency of the calibrator signal. May be unspecified.
\item [] \texttt{CAL\_MODE}: Setting of calibration noise diode
\item [] \texttt{CHAN\_BW}: Bandwidth (in MHz) of a single channel.  Will be negative if the frequency axis is reversed (e.g.\ due to a lower sideband mix).
\item [] \texttt{CHAN\_DM}: Dispersion measure for COHERENT\_SEARCH mode. Set to 0 for all other modes.
\item [] \texttt{DAQCTRL}: This field is used by the telescope interface code to control the data acquisition (e.g.\ to initiate recording).
\item [] \texttt{DAQPULSE}: Heartbeat from GUPPI DAQ (data acquisition) server
\item [] \texttt{DAQSTATE}: State of GUPPI DAQ server
\item [] \texttt{DATADIR}: Path prefix for data directory of output files
\item [] \texttt{DATAHOST}: Hostname or IP address of the FPGA 10Gbe port to receive from
\item [] \texttt{DATAPORT}: Port number for receiving data
\item [] \texttt{DEC\_STR}: Declination of the telescope (string form)
\item [] \texttt{DEC}: Declination of the telescope (degrees) (Parkes: same as DEC\_STR)
\item [] \texttt{DIRECTIO}: Indicates whether Direct I/O was used to optimize file writing. If used, indicated by a non-zero value, the header is padded to be a multiple of 512 bytes.
\item [] \texttt{DISKSTAT}: Status of disk write thread
\item [] \texttt{DROPAVG}: Average fraction of recently dropped data
\item [] \texttt{DROPBLK}: Fraction of data dropped in the current block
\item [] \texttt{DROPTOT}: Fraction of total data dropped during this scan
\item [] \texttt{DS\_FREQ}: Enables software reduction of frequency resolution (SEARCH and COHERENT\_SEARCH only). Only powers of 2 are allowed. Set to 1 for all other modes.
\item [] \texttt{DS\_TIME}: Enables additional software accumulation of spectra in time (SEARCH and COHERENT\_SEARCH only). Only powers of 2 are allowed. Set to 1 for all other modes.
\item [] \texttt{FD\_POLN}: Polarization type of the feed (LIN or CIR)
\item [] \texttt{FFTLEN}: Total number of frequency channels in the digitized band from which these data are drawn.  Not used by Breakthrough Listen code.
\item [] \texttt{FREQ}: Parkes: center frequency tuning according to TCS (MHz)
\item [] \texttt{FRONTEND}: Name of the receiver that was used. 
\item [] \texttt{LST}: Local sidereal time at the start of the block (seconds)
\item [] \texttt{NBIN}: Number of bins in profile for folding and calibration modes. Ignored in all other modes.
\item [] \texttt{NBITS}: Number of bits in each complex component per sample -- one complex-valued sample has $2 \times \mathrm{NBITS}$ bits
\item [] \texttt{NDROP}: Number of packets dropped during this block
\item [] \texttt{NETBUFST}: Recording diagnostic regarding how many network buffers were used during recording -- usually only 1 out of 24 are required, but if there are disk or filesystem bottlenecks this goes up.
\item [] \texttt{NETSTAT}: Status of network thread
\item [] \texttt{NPKT}: Number of packets received for this block
\item [] \texttt{NPOL}: Number of samples per time step -- 4 corresponds to dual-polarization complex data
\item [] \texttt{NRCVR}: Number of receiver channels (polarizations)
\item [] \texttt{OBSBW}: Width of passband in this node; negative sign indicates reversed frequency axis
\item [] \texttt{OBSERVER}: The observer on duty when these data were recorded
\item [] \texttt{OBSFREQ}: Center frequency of the frequencies spanned by the data in the RAW file
\item [] \texttt{OBSNCHAN}: The number of frequency channels contained within the file.
\item [] \texttt{OBS\_MODE}: Current observing mode (SEARCH, FOLD, CAL, COHERENT\_SEARCH, COHERENT\_FOLD, COHERENT\_CAL, or RAW)
\item [] \texttt{OFFSET0}: Not used
\item [] \texttt{OFFSET1}: Not used
\item [] \texttt{OFFSET2}: Not used
\item [] \texttt{OFFSET3}: Not used
\item [] \texttt{ONLY\_I}: Setting this to 1 enables total intensity data to be recorded in SEARCH and COHERENT\_SEARCH only. Otherwise, this will be set to 0 for full-Stokes data.
\item [] \texttt{OVERLAP}: Number of time samples per channel overlapping between blocks.  Always 0 for Breakthrough Listen data.
\item [] \texttt{PARFILE}: For pulsar folding modes, full path the parameter file. Ignored in other modes.
\item [] \texttt{PFB\_OVER}: PFB number of overlapping channels.  Not used by Breakthrough Listen code.
\item [] \texttt{PKTFMT}: UDP packet format. FAST4K or 1SFA.
\item [] \texttt{PKTIDX}: Recording diagnostic -- current first packet ID \# of this block
\item [] \texttt{PKTSIZE}: Packet size (bytes)
\item [] \texttt{PKTSTART}: This is the expected PKTIDX value at the start of the scan (i.e.\ when recording should start).
\item [] \texttt{PKTSTOP}: This is the PKTIDX value at the end of the scan (i.e.\ when recording should stop).
\item [] \texttt{POL\_TYPE}: Not used
\item [] \texttt{PROJID}: Observing project ID / session code
\item [] \texttt{RA\_STR}: Right Ascension of the telescope (string form)
\item [] \texttt{RA}: Right ascension of the telescope (degrees) (Parkes: same as RA\_STR)
\item [] \texttt{SCALE0}: Not used
\item [] \texttt{SCALE1}: Not used
\item [] \texttt{SCALE2}: Not used
\item [] \texttt{SCALE3}: Not used
\item [] \texttt{SCANLEN}: Overall duration of the scan (i.e.\ recording) in seconds.
\item [] \texttt{SCANREM}: Seconds of scan time remaining.  Intended primarily as feedback for the observer while a scan is occurring.
\item [] \texttt{SCAN}: GBT scan index \#
\item [] \texttt{SRC\_NAME}: Name of current target, taken from the input catalog.
\item [] \texttt{STTVALID}: 1 if valid start time has been set
\item [] \texttt{STT\_IMJD}: Start time, integer MJD
\item [] \texttt{STT\_OFFS}: Start time, fractional second
\item [] \texttt{STT\_SMJD}: Start time, integer seconds within day (UTC)
\item [] \texttt{SYNCTIME}: (Parkes) Current UNIX epoch seconds.
\item [] \texttt{TBIN}: Time resolution of incoming data, (1 / abs(CHAN\_BW))  
\item [] \texttt{TELESCOP}: Telescope name
\item [] \texttt{TFOLD}: Not used
\item [] \texttt{TRK\_MODE}: Tracking mode in use
\item [] \texttt{ZA}: Zenith Angle of the telescope 
\end{itemize}

\clearpage

\begin{figure}[ht]
\centering
\twocolumngrid
\input{rawheader.tex}
    \caption{Typical raw header for BL data}
    \label{rawheader}
\end{figure}

\subsection{Filterbank/HDF5 Metadata}

The metadata within these reduced file types contain deprecated or unused fields. The most important, noting source, time, location, and shape of the data payload, are listed in Table~\ref{tab:watutiloutput}.

\begin{deluxetable*}{l r l}
\tablewidth{0pt}
\tabletypesize{\scriptsize}
\tablecaption{\label{tab:watutiloutput} A subset of metadata from a filterbank file, as presented by the \texttt{watutil} command}
\tablehead{
\colhead{Key} &
\colhead{Example Value} &
\colhead{Description}
}
\startdata
     rawdatafile & guppi\_58598\_54807\_HIP112437\_0158.0000.raw & Original raw data input \\
     source\_name &                        HIP112437 & Target/source name \\
          tstart &                    58598.6343403 & MJD of start of scan \\
          nchans &                         67108864 & Number of frequency channels \\
           tsamp &                     18.253611008 & Length of sample integration (secs) \\
            foff &               -2.79396772385e-06 & Size of each frequency channel \\
         src\_raj &                     22:46:33.408 & Right ascension of source (hours)\\
         src\_dej &                      59:40:26.76 & Declination of source (deg) \\
            fch1 &                    11251.4648424 & Frequency of first channel in file \\
           nbits &                               32 & Number of bits per channel \\
\enddata

\end{deluxetable*}


\section{Frequency Tunings and Data Products}

\begin{rotatetable}
\begin{deluxetable*}{l c c c c c c c c}
\tablewidth{0pt}
\tabletypesize{\scriptsize}
\tablecaption{\label{tab:freqs} GBT band frequency table.}
\tablehead {
\nocolhead{} &
\twocolhead{L-band} &
\twocolhead{S-band} &
\twocolhead{C-band} &
\twocolhead{X-band} \\
\nocolhead{} &
\twocolhead{(1.1--1.9\,GHz)} &
\twocolhead{(1.7--2.7\,GHz)} &
\twocolhead{(4.0--8.0\,GHz)} &
\twocolhead{(8.0--11.6\,GHz)} \\
\colhead{Node} &
\colhead{.raw {\sc obsfreq}} &
\colhead{.fil fch1} &
\colhead{.raw {\sc obsfreq}} &
\colhead{.fil fch1} &
\colhead{.raw {\sc obsfreq}} &
\colhead{.fil fch1} &
\colhead{.raw {\sc obsfreq}} &
\colhead{.fil fch1}
}
\startdata
blc00 & 2157.71484375 & 2251.46484375 & 3057.71484375 & 3151.46484375 & 8345.21484375 & 8438.96484375 & 11157.71484375 & 11251.4648437 \\
blc01 & 1970.21484375 & 2063.96484375 & 2870.21484375 & 2963.96484375 & 8157.71484375 & 8251.46484375 & 10970.21484375 & 11063.9648437 \\
blc02 & 1782.71484375 & 1876.46484375 & 2682.71484375 & 2776.46484375 & 7970.21484375 & 8063.96484375 & 10782.71484375 & 10876.4648437 \\
blc03 & 1595.21484375 & 1688.96484375 & 2495.21484375 & 2588.96484375 & 7782.71484375 & 7876.46484375 & 10595.21484375 & 10688.9648437 \\
blc04 & 1407.71484375 & 1501.46484375 & 2307.71484375 & 2401.46484375 & 7595.21484375 & 7688.96484375 & 10407.71484375 & 10501.4648437 \\
blc05 & 1220.21484375 & 1313.96484375 & 2120.21484375 & 2213.96484375 & 7407.71484375 & 7501.46484375 & 10220.21484375 & 10313.9648437 \\
blc06 & 1032.71484375 & 1126.46484375 & 1932.71484375 & 2026.46484375 & 7220.21484375 & 7313.96484375 & 10032.71484375 & 10126.4648437 \\
blc07 &  845.21484375 &  938.96484375 & 1745.21484375 & 1838.96484375 & 7032.71484375 & 7126.46484375 & 9845.21484375  & 9938.96484375 \\
blc10 &               &               &               &               & 7220.21484375 & 7313.96484375 & 10032.71484375 & 10126.4648437 \\
blc11 &               &               &               &               & 7032.71484375 & 7126.46484375 & 9845.21484375  & 9938.96484375 \\
blc12 &               &               &               &               & 6845.21484375 & 6938.96484375 & 9657.71484375  & 9751.46484375 \\
blc13 &               &               &               &               & 6657.71484375 & 6751.46484375 & 9470.21484375  & 9563.96484375 \\
blc14 &               &               &               &               & 6470.21484375 & 6563.96484375 & 9282.71484375  & 9376.46484375 \\
blc15 &               &               &               &               & 6282.71484375 & 6376.46484375 & 9095.21484375  & 9188.96484375 \\
blc16 &               &               &               &               & 6095.21484375 & 6188.96484375 & 8907.71484375  & 9001.46484375 \\
blc17 &               &               &               &               & 5907.71484375 & 6001.46484375 & 8720.21484375  & 8813.96484375 \\
blc20 &               &               &               &               & 6095.21484375 & 6188.96484375 & 8907.71484375  & 9001.46484375 \\
blc21 &               &               &               &               & 5907.71484375 & 6001.46484375 & 8720.21484375  & 8813.96484375 \\
blc22 &               &               &               &               & 5720.21484375 & 5813.96484375 & 8532.71484375  & 8626.46484375 \\
blc23 &               &               &               &               & 5532.71484375 & 5626.46484375 & 8345.21484375  & 8438.96484375 \\
blc24 &               &               &               &               & 5345.21484375 & 5438.96484375 & 8157.71484375  & 8251.46484375 \\
blc25 &               &               &               &               & 5157.71484375 & 5251.46484375 & 7970.21484375  & 8063.96484375 \\
blc26 &               &               &               &               & 4970.21484375 & 5063.96484375 & 7782.71484375  & 7876.46484375 \\
blc27 &               &               &               &               & 4782.71484375 & 4876.46484375 & 7595.21484375  & 7688.96484375 \\
blc30 &               &               &               &               & 4970.21484375 & 5063.96484375 &                &               \\
blc31 &               &               &               &               & 4782.71484375 & 4876.46484375 &                &               \\
blc32 &               &               &               &               & 4595.21484375 & 4688.96484375 &                &               \\
blc33 &               &               &               &               & 4407.71484375 & 4501.46484375 &                &               \\
blc34 &               &               &               &               & 4220.21484375 & 4313.96484375 &                &               \\
blc35 &               &               &               &               & 4032.71484375 & 4126.46484375 &                &               \\
blc36 &               &               &               &               & 3845.21484375 & 3938.96484375 &                &               \\
blc37 &               &               &               &               & 3657.71484375 & 3751.46484375 &                &               \\
\enddata
\end{deluxetable*}
\end{rotatetable}

\begin{deluxetable*}{l c c | l c c}
\tablewidth{0pt}
\tabletypesize{\scriptsize}
\tablecaption{\label{tab:pksfreqs} Parkes Telescope 10CM band frequency table. The ``X" in the 10CM node name should be replaced by 0, 1, or 2 depending on which bank of nodes recorded that session. ``AB" in the Multibeam node name refers to any given beam pair, where beam = $\mathrm{int}(((A \times 8)+B-1)/2)$; ``AC" is the next highest node. For example, blc01 and blc02 are beam 0, blc03 and blc04 are beam 1, blc05 and blc06 are beam 2, blc07 and blc10 are beam 3, etc.}
\tablehead {
\nocolhead{} &
\twocolhead{10CM} &
\nocolhead{} &
\twocolhead{Multibeam} \\
\nocolhead{} &
\twocolhead{(2.7--3.5\,GHz)} &
\nocolhead{} &
\twocolhead{(1.2--1.5\,GHz)} \\
\colhead{Node} &
\colhead{.raw {\sc obsfreq}} &
\colhead{.fil fch1} &
\colhead{Node} &
\colhead{.raw {\sc obsfreq}} &
\colhead{.fil fch1} \\
}
\startdata
blcX0 & 2667.78515625 & 2574.03515625 & blcAB & 1438.5 & 1513.75 \\
blcX1 & 2855.28515625 & 2761.53515625 & blcAC & 1284.5 & 1359.75 \\
blcX2 & 3042.78515625 & 2949.03515625 & & &  \\
blcX3 & 3230.28515625 & 3136.53515625 & & &  \\
blcX4 & 3417.78515625 & 3324.03515625 & & &  \\
blcX5 & 3605.28515625 & 3511.53515625 & & &  \\
\enddata
\end{deluxetable*}

\clearpage

\section{Summary of Available Data Products}\label{app:summary}

At the time of writing, there are 94,534 files available for public download, described in more detail in Table~\ref{tab:datasummary}.

\begin{deluxetable*}{l c c c c r}
\tablewidth{0pt}
\tabletypesize{\scriptsize}
\tablecaption{\label{tab:datasummary} Number of publicly available data files separated by project and file type.}
\tablehead{
\nocolhead{} &
\multicolumn{4}{c}{Format} &
\nocolhead{} \\
\colhead{Project} & 
\colhead{FITS} &
\colhead{raw voltage} & 
\colhead{filterbank} & 
\colhead{HDF5} & 
\colhead{Total} \\
}
\startdata
APF & 6086 & & & & 6086 \\
GBT & & 40375 & 1456 & 31094 & 72925 \\
Parkes & & 7358 & & 8165 & 15523 \\
\enddata
\end{deluxetable*}

Note: In this data release all Green Bank data products are in descending frequency order, and Parkes products are in ascending frequency.

Within these files are several important subsets, easily findable via searches on the open data web sites noted in Section \ref{archive}:

\begin{itemize}
\item The FRB121102 repeater data set: this contains 24608 raw voltage and 1066 reduced filterbank files (each scan of about 70 raw data files reduces into the three standard 3 filterbank files, hence the disparity in amounts). This data set is presented unspliced - each data product represents 187.5MHz of the total bandwidth.
\item HDF5 data for everything analyzed in \citet{enriquez2017} and \citePrice{}: all the HFD5 data currently available are fine frequency resolution products used in the SETI analysis of these papers, as well as the other standard resolutions. 
\item APF FITS image data: all the data of this format are from the entire Automated Planet Finder survey, and is continually growing. 
\end{itemize}

A list of the larger/notable data sets available in this data release are listed in Table~\ref{tab:targetsummary}. You can find data from any of these catagories by searching for target names that begin with the listed substrings. For example, all the Hipparcos star catalog targets begin with the substring  ``HIP." 

\begin{deluxetable*}{l l c c c c}
\tablewidth{0pt}
\tabletypesize{\scriptsize}
\tablecaption{\label{tab:targetsummary} Larger/notable data sets available, sorted by the first characters in the target name (noted for easier searching).}
\tablehead{
\nocolhead{} &
\colhead{Target names} &
\multicolumn{4}{c}{Format} \\
\colhead{Data set} &
\colhead{starting with} & 
\colhead{FITS} &
\colhead{raw voltage} & 
\colhead{filterbank} & 
\colhead{HDF5} \\
}
\startdata
Alpha Centuri & ALPHACEN & 0 & 0 & 0 & 191 \\
FRB121102 (repeating FRB) & FRB121102 & 0 & 24608 & 1066 & 0 \\
Gliese star catalog & GJ & 186 & 0 & 0 & 912 \\
Hipparcos star catalog & HIP & 5528 & 10002 & 0 & 36217 \\
Various pulsars & J / PSR\_J & 0 & 1823 & 192 & 281 \\
LP944-020 source & LP944-020 & 0 & 0 & 0 & 173 \\
Andromeda galaxy & MESSIER031 & 0 & 3156 & 0 & 0 \\
'Oumuamua & OUMUAMUA & 0 & 8 & 6 & 0 \\
Proxima Centuri & PROXCEN & 0 & 0 & 0 & 222 \\
Voyager I & VOYAGER1 & 0 & 670 & 0 & 0 \\
\enddata
\end{deluxetable*}

\clearpage

\bibliographystyle{aasjournal}
\bibliography{references}

\begin{thebibliography}{}
\expandafter\ifx\csname natexlab\endcsname\relax\def\natexlab#1{#1}\fi
\providecommand{\url}[1]{\href{#1}{#1}}

\bibitem[{Ini(2016)}]{Initiatives:2016up}
 2016, {National Astronomical Observatories of China, Breakthrough Initiatives
  Launch Global Collaboration in Search for Intelligent Life in the Universe},
  , .
\newblock \url{https://breakthroughinitiatives.org/news/6}

\bibitem[{Anderson {et~al.}(2002)Anderson, Cobb, Korpela, Lebofsky, \&
  Werthimer}]{Anderson:2002:SEP:581571.581573}
Anderson, D.~P., Cobb, J., Korpela, E., Lebofsky, M., \& Werthimer, D. 2002,
  Commun. ACM, 45, 56.
\newblock \url{http://doi.acm.org/10.1145/581571.581573}

\bibitem[{{BL Collaboration}(2019)}]{ascl:blimpy}
{BL Collaboration}. 2019, Astrophysics Source Code Library, ascl:submitted

\bibitem[{Croft {et~al.}(2016)Croft, Siemion, MacMahon, Lebofsky, Hickish,
  Price, Deboer, Werthimer, Gajjar, \& Isaacson}]{Croft:2016tm}
Croft, S., Siemion, A., MacMahon, D., {et~al.} 2016, researchgate.net

\bibitem[{Enriquez {et~al.}(2017)Enriquez, Siemion, Foster, Gajjar, Hellbourg,
  Hickish, Isaacson, Price, Croft, DeBoer, Lebofsky, MacMahon, \&
  Werthimer}]{enriquez2017}
Enriquez, J.~E., Siemion, A., Foster, G., {et~al.} 2017, ApJ, 849, 104.
\newblock
  \url{http://adsabs.harvard.edu/cgi-bin/nph-data_query?bibcode=2017ApJ...849..104E&link_type=EJOURNAL}

\bibitem[{{Enriquez} {et~al.}(2018){Enriquez}, {Siemion}, {Lazio}, {Lebofsky},
  {MacMahon}, {Park}, {Croft}, {DeBoer}, {Gizani}, {Gajjar}, {Hellbourg},
  {Isaacson}, \& {Price}}]{2018RNAAS...2a...9E}
{Enriquez}, J.~E., {Siemion}, A., {Lazio}, T. J.~W., {et~al.} 2018, Research
  Notes of the American Astronomical Society, 2, 9

\bibitem[{{Enriquez} {et~al.}(2019){Enriquez}, {Siemion}, {Dana}, {Croft},
  {M{\'e}ndez}, {Xu}, {Deboer}, {Gajjar}, {Hellbourg}, {Isaacson}, {Lebofsky},
  {MacMahon}, {Price}, {Werthimer}, \& {Zuluaga}}]{Enriquez2019}
{Enriquez}, J.~E., {Siemion}, A., {Dana}, R., {et~al.} 2019, International
  Journal of Astrobiology, 18, 33

\bibitem[{{Gajjar} {et~al.}(2018){Gajjar}, {Siemion}, {Price}, {Law},
  {Michilli}, {Hessels}, {Chatterjee}, {Archibald}, {Bower}, {Brinkman},
  {Burke-Spolaor}, {Cordes}, {Croft}, {Enriquez}, {Foster}, {Gizani},
  {Hellbourg}, {Isaacson}, {Kaspi}, {Lazio}, {Lebofsky}, {Lynch}, {MacMahon},
  {McLaughlin}, {Ransom}, {Scholz}, {Seymour}, {Spitler}, {Tendulkar},
  {Werthimer}, \& {Zhang}}]{gajjar:18}
{Gajjar}, V., {Siemion}, A.~P.~V., {Price}, D.~C., {et~al.} 2018, \apj, 863, 2

\bibitem[{{Holden} {et~al.}(2016){Holden}, {Burt}, \& {Deich}}]{Holden2016}
{Holden}, B.~P., {Burt}, J.~A., \& {Deich}, W. T.~S. 2016, in Society of
  Photo-Optical Instrumentation Engineers (SPIE) Conference Series, Vol. 9910,
  Observatory Operations: Strategies, Processes, and Systems VI, 99102A

\bibitem[{Isaacson {et~al.}(2017)Isaacson, Siemion, Marcy, Lebofsky, Price,
  MacMahon, Croft, DeBoer, Hickish, Werthimer, Sheikh, Hellbourg, \&
  Enriquez}]{Isaacson2017}
Isaacson, H., Siemion, A. P.~V., Marcy, G.~W., {et~al.} 2017, arXiv,
  arXiv:1701.06227.
\newblock \url{http://adsabs.harvard.edu/abs/2017arXiv170106227I}

\bibitem[{{Lipman} {et~al.}(2019){Lipman}, {Isaacson}, {Siemion}, {Lebofsky},
  {Price}, {MacMahon}, {Croft}, {DeBoer}, {Hickish}, {Werthimer}, {Hellbourg},
  {Enriquez}, \& {Gizani}}]{lipman:19}
{Lipman}, D., {Isaacson}, H., {Siemion}, A.~P.~V., {et~al.} 2019, \pasp, 131,
  034202

\bibitem[{Lorimer(2006)}]{Lorimer:2006vv}
Lorimer, D.~R. 2006, {SIGPROC{\textendash}v3.7 : (Pulsar) Signal Processing
  Programs } (sigproc.sourceforge.net)

\bibitem[{{MacMahon} {et~al.}(2018){MacMahon}, {Price}, {Lebofsky}, {Siemion},
  {Croft}, {DeBoer}, {Enriquez}, {Gajjar}, {Hellbourg}, {Isaacson},
  {Werthimer}, {Abdurashidova}, {Bloss}, {Brandt}, {Creager}, {Ford}, {Lynch},
  {Maddalena}, {McCullough}, {Ray}, {Whitehead}, \& {Woody}}]{gbtinstrument}
{MacMahon}, D.~H.~E., {Price}, D.~C., {Lebofsky}, M., {et~al.} 2018, \pasp,
  130, 044502

\bibitem[{{Margot} {et~al.}(2018){Margot}, {Greenberg}, {Pinchuk}, {Shinde},
  {Alladi}, {Prasad MN}, {Bowman}, {Fisher}, {Gyalay}, {McKibbin}, {Miles},
  {Nguyen}, {Power}, {Ramani}, {Raviprasad}, {Santana}, \&
  {Lynch}}]{Margot:2018}
{Margot}, J.-L., {Greenberg}, A.~H., {Pinchuk}, P., {et~al.} 2018, \aj, 155,
  209

\bibitem[{{Pence} {et~al.}(2010){Pence}, {Chiappetti}, {Page}, {Shaw}, \&
  {Stobie}}]{fits3}
{Pence}, W.~D., {Chiappetti}, L., {Page}, C.~G., {Shaw}, R.~A., \& {Stobie}, E.
  2010, \aap, 524, A42

\bibitem[{{Pinchuk} {et~al.}(2019){Pinchuk}, {Margot}, {Greenberg}, {Ayalde},
  {Bloxham}, {Boddu}, {Gerardo Chinchilla-Garcia}, {Cliffe}, {Gallagher},
  {Hart}, {Hesford}, {Mizrahi}, {Pike}, {Rodger}, {Sayki}, {Schneck}, {Tan},
  {{\textquotedblleft}Yolanda{\textquotedblright} Xiao}, \&
  {Lynch}}]{Pinchuk:2019}
{Pinchuk}, P., {Margot}, J.-L., {Greenberg}, A.~H., {et~al.} 2019, \aj, 157,
  122

\bibitem[{{Price}(2016)}]{2016arXiv160703579P}
{Price}, D.~C. 2016, arXiv e-prints, arXiv:1607.03579

\bibitem[{{Price} {et~al.}(2018){Price}, {MacMahon}, {Lebofsky}, {Croft},
  {DeBoer}, {Enriquez}, {Foster}, {Gajjar}, {Gizani}, {Hellbourg}, {Isaacson},
  {Siemion}, {Werthimer}, {Green}, {Amy}, {Ball}, {Bock}, {Craig}, {Edwards},
  {Jameson}, {Mader}, {Preisig}, {Smith}, {Reynolds}, \&
  {Sarkissian}}]{parkesinstrument}
{Price}, D.~C., {MacMahon}, D.~H.~E., {Lebofsky}, M., {et~al.} 2018, \pasa, 35,
  arXiv:1804.04571

\bibitem[{{Price} {et~al.}(2019){Price}, {Croft}, {DeBoer}, {Drew}, {Enriquez},
  {Foster}, {Gajjar}, {Gizani}, {Hellbourg}, {Isaacson}, {Lebofsky},
  {MacMahon}, {de Pater}, {Siemion}, {Worden}, \&
  {Zhang}}]{2019RNAAS...3a..19P}
{Price}, D.~C., {Croft}, S., {DeBoer}, D., {et~al.} 2019, Research Notes of the
  American Astronomical Society, 3, 19

\bibitem[{{Radovan} {et~al.}(2014){Radovan}, {Lanclos}, {Holden}, {Kibrick},
  {Allen}, {Deich}, {Rivera}, {Burt}, {Fulton}, {Butler}, \&
  {Vogt}}]{Radovan2014}
{Radovan}, M.~V., {Lanclos}, K., {Holden}, B.~P., {et~al.} 2014, in \procspie,
  Vol. 9145, Ground-based and Airborne Telescopes V, 91452B

\bibitem[{{Ransom}(2011)}]{presto}
{Ransom}, S. 2011, {PRESTO: PulsaR Exploration and Search TOolkit}, , ,
  ascl:1107.017

\bibitem[{{Ransom} {et~al.}(2009){Ransom}, {Demorest}, {Ford}, {McCullough},
  {Ray}, {DuPlain}, \& {Brandt}}]{guppi}
{Ransom}, S., {Demorest}, P., {Ford}, J., {et~al.} 2009, 214

\bibitem[{Siemion {et~al.}(2013)Siemion, Demorest, Korpela, Maddalena,
  Werthimer, Cobb, Howard, Langston, Lebofsky, Marcy, \&
  Tarter}]{2013ApJ...767...94S}
Siemion, A. P.~V., Demorest, P., Korpela, E., {et~al.} 2013, The Astrophysical
  Journal, 767, 94

\bibitem[{{Tellis} \& {Marcy}(2015)}]{Tellis2015}
{Tellis}, N.~K., \& {Marcy}, G.~W. 2015, \pasp, 127, 540

\bibitem[{{Tellis} \& {Marcy}(2017)}]{Tellis2017}
---. 2017, \aj, 153, 251

\bibitem[{Thompson {et~al.}(1986)Thompson, Moran, Swenson,
  {et~al.}}]{thompson1986interferometry}
Thompson, A.~R., Moran, J.~M., Swenson, G.~W., {et~al.} 1986, Interferometry
  and synthesis in radio astronomy (Wiley New York et al.)

\bibitem[{Tingay {et~al.}(2018)Tingay, Tremblay, \&
  Croft}]{2018ApJ...856...31T}
Tingay, S.~J., Tremblay, C.~D., \& Croft, S. 2018, The Astrophysical Journal,
  856, 31

\bibitem[{{van Straten} \& {Bailes}(2011)}]{2011PASA...28....1V}
{van Straten}, W., \& {Bailes}, M. 2011, Publications of the Astronomical
  Society of Australia, 28, 1

\bibitem[{{van Straten} {et~al.}(2012){van Straten}, {Demorest}, \&
  {Oslowski}}]{2012AR&T....9..237V}
{van Straten}, W., {Demorest}, P., \& {Oslowski}, S. 2012, Astronomical
  Research and Technology, 9, 237

\bibitem[{{Wells} {et~al.}(1981){Wells}, {Greisen}, \& {Harten}}]{fits}
{Wells}, D.~C., {Greisen}, E.~W., \& {Harten}, R.~H. 1981, \aaps, 44, 363

\bibitem[{{West} \& {Hilburn}(2019)}]{West_19:fosdem}
{West}, N., \& {Hilburn}, B. 2019, libsigmf: Human Tools for Extra-Terrestrial
  and AI Radios, Talk at FOSDEM'19, Brussels, ,

\bibitem[{Worden {et~al.}(2018)Worden, Drew, \&
  Klupar}]{doi:10.1089/space.2018.0027}
Worden, S.~P., Drew, J., \& Klupar, P. 2018, New Space, 6, 262.
\newblock \url{https://doi.org/10.1089/space.2018.0027}

\bibitem[{{Wright} \& {Sigurdsson}(2016)}]{Wright2016}
{Wright}, J.~T., \& {Sigurdsson}, S. 2016, \apj, 829, L3

\bibitem[{{Zhang} {et~al.}(2018){Zhang}, {Gajjar}, {Foster}, {Siemion},
  {Cordes}, {Law}, \& {Wang}}]{zhang:18}
{Zhang}, Y.~G., {Gajjar}, V., {Foster}, G., {et~al.} 2018, \apj, 866, 149

\end{thebibliography}

\end{document}